# Deregulation of calcium homeostasis in Bcr-Abl-dependent chronic myeloid leukemia

**Hélène Cabanas[1], Thomas Harnois[1], Christophe Magaud[2], Laëtitia Cousin[1], Bruno Constantin[1], Nicolas Bourmeyster[1] and Nadine Déliot[1]**

[1]Laboratoire de Signalisation et Transports Ioniques Membranaires (STIM) ERL CNRS 7368, Equipe Calcium et Microenvironnement des Cellules Souches (CMCS), Université de Poitiers, 86073 Poitiers, France

[2]Laboratoire de Signalisation et Transports Ioniques Membranaires (STIM) ERL CNRS 7368, Equipe Transferts Ioniques et Rythmicité Cardiaque (TIRC), Université de Poitiers, 86073 Poitiers, France

*Correspondence to:* Bruno Constantin, *email:* bruno.constantin@univ-poitiers.fr

## ABSTRACT

**Background:** Chronic myeloid leukemia (CML) results from hematopoietic stem cell transformation by the *bcr-abl* chimeric oncogene, encoding a 210 kDa protein with constitutive tyrosine kinase activity. In spite of the efficiency of tyrosine kinase inhibitors (TKI; Imatinib), other strategies are explored to eliminate CML leukemia stem cells, such as calcium pathways.

**Results:** In this work, we showed that Store-Operated Calcium Entry (SOCE) and thrombin induced calcium influx were decreased in Bcr-Abl expressing 32d cells (32d-p210). The 32d-p210 cells showed modified Orai1/STIM1 ratio and reduced TRPC1 expression that could explain SOCE reduction. Decrease in SOCE and thrombin induced calcium entry was associated to reduced Nuclear Factor of Activated T cells (NFAT) nucleus translocation in 32d-p210 cells. We demonstrated that SOCE blockers enhanced cell mobility of 32d-p210 cells and reduced the proliferation rate in both 32d cell lines. TKI treatment slightly reduced the thrombin-induced response, but imatinib restored SOCE to the wild type level. Bcr-Abl is also known to deregulate Protein Kinase C (PKC), which was described to modulate calcium entries. We showed that PKC enhances SOCE and thrombin induced calcium entries in control cells while this effect is lost in Bcr-Abl-expressing cells.

**Conclusion:** The tyrosine kinase activity seems to regulate calcium entries probably not directly but through a global cellular reorganization involving a PKC pathway. Altogether, calcium entries are deregulated in Bcr-Abl-expressing cells and could represent an interesting therapeutic target in combination with TKI.

## INTRODUCTION

Chronic myeloid leukemia (CML) results from hematopoietic stem cell transformation by the *bcr-abl* chimeric oncogene, generated by a reciprocal translocation between chromosomes 9 and 22 (Philadelphia chromosome, Ph+) [1]. CML is a myeloproliferative disorder that progresses from initial chronic phase to accelerated phase and terminal blast crisis. The structure of the generated protein (p210[bcr-abl]) consists of functional domains which are successively oligomerization, serine/threonine kinase, RhoA-specific DH/PH (Dbl-Homology/Pleckstrin homology), SH2 and SH3, tyrosine kinase and actin binding domains [2]. In parallel to the tyrosine kinase domain, responsible for leukemogenesis, all these domains participate to the general deregulation of the intracellular signaling pathways in Bcr-Abl-expressing cells. Bcr-Abl tyrosine kinase inhibitors (TKI) are remarkably effective in inducing remissions, preventing transformation to blast crisis, and prolonging survival of CML patients [3]. However, TKI treatment fails to eliminate leukemic stem cells (LSC), even in patients achieving deep molecular responses [4]. Although a subset of CML chronic phase patients are able to maintain remission after stopping TKI, most patients require continuous treatment to prevent relapse [5]. Therefore, improved understanding of the mechanisms of LSC resistance to TKI should help to develop new strategies to obtain treatment-free remissions.

In CML, the constitutively activated tyrosine-kinase activity of Bcr-Abl generates in stem cells, but also in differentiated cells, multiple signaling abnormalities, leading to enhanced proliferation, inhibited apoptosis and stimulated migration. It prompted numerous research teams to explore other signaling pathways that could provide alternate therapeutic targets, especially for treating cancer progenitors. Calcium ($Ca^{2+}$) signaling pathways could be an alternate therapeutic target because of its importance in various cellular processes such as proliferation, protein expression and migration in normal and cancer cells. In particular, $Ca^{2+}$ channels expressed at the plasma membrane (PM) could offer potential therapeutic targets and/or prognostic markers. $Ca^{2+}$ channels are indeed major actors that transduce microenvironment signals into spatio-temporal fluctuations of intracellular $Ca^{2+}$ concentrations and in turn modulate specific cellular mechanisms. The amplitude of the cytoplasmic $Ca^{2+}$ variation regulates the responses such as local or global $Ca^{2+}$ increase coming from extracellular entries and/or intracellular stock release [6]. Cancer cells hijack the physiological functions of those channels, which then contribute to the neoplastic phenotype.

In non-excitable cells, the main $Ca^{2+}$ entry is mediated through the binding of an agonist (thrombin, catecholamine, bradykinin, etc.) on its tyrosine kinase receptor or G protein coupled receptor (GPCR) which activates phospholipase C (PLC). PLC mediates the cleavage of phosphatidylinositol 4,5-bisphosphate ($PIP_2$) into diacylglycerol (DAG) and inositol 1,4,5-trisphosphate ($IP_3$). DAG stimulates receptor-operated channels (ROCs) via protein kinase C (PKC) activation. ROCs are mainly members of transient receptor potential canonical (TRPC) family (for example TRPC1-3-6) and contribute to the cytosolic $Ca^{2+}$ increase [7]. In parallel to PKC activation, the formation of $IP_3$ triggers $Ca^{2+}$ release from the endoplasmic reticulum (ER) which stimulates store-operated channels (SOCs) (for review: [8, 9]). Stromal interaction molecule 1 (STIM1) is an ER $Ca^{2+}$ sensor that oligomerizes when ER $Ca^{2+}$ concentration diminishes, moves to the ER/PM junctions and then binds to the PM Orai1 channels [10–12]. STIM1 and Orai1 are the main components of store-operated $Ca^{2+}$ entry (SOCE), also called $Ca^{2+}$ release-activated $Ca^{2+}$ (CRAC) channels. The molecular interaction of STIM1/Orai is now well documented [13–16]. Moreover, different members of TRPC family have also been involved in SOCE (for review: [17]). For example, TRPC1 can interact with STIM1 [18, 19] and in turn contribute to SOCE in different tissues (salivary gland, smooth muscle, endothelial, etc.) (for review: [7, 20]). In salivary gland cells, STIM1/Orai1 dependent $Ca^{2+}$ entry contributes to the targeting of TRPC1 from vesicles to PM in the ER/PM junctions. TRPC1 translocation allows the co-localization with STIM1 and Orai1 and then TRPC1 participates to SOCE amplification [21, 22].

Because $Ca^{2+}$ is involved in numerous cellular processes, $Ca^{2+}$ regulation has also been studied in various pathologies and especially in cancer. $Ca^{2+}$ channels and pumps are known to be deregulated in various tumors and could be used as cancer hallmarks (for review: [23–25]). STIM, Orai and TRP channels and in turn $Ca^{2+}$ homeostasis play a role in cell processes that are modified in tumor context [26–30] but their expression levels and activation are dependent of cancers types and will have different functional consequences.

In CML, very few studies investigate the link between $Ca^{2+}$ homeostasis and leukemogenesis. In fact, the two publications exploring the relationships between Bcr-Abl expression and $Ca^{2+}$ are now relatively old and propose opposite results [31, 32]. In Bcr-Abl-expressing cells, Piwocka *et al*., showed that the Bcr-Abl-expressing cells have a decrease in ER $Ca^{2+}$ release as well as a reduced SOCE, which inhibit $Ca^{2+}$-dependent apoptosis [31]. In another study, the tyrosine kinase activity of Bcr-Abl enhances SOCE via PKC inhibition, that positively regulates survival factors [32]. These works were done concurrently with the discovery of the molecular actors of SOCE: Orai/STIM. Therefore, few data are available for $Ca^{2+}$/SOCE/CML and the current treatment of CML inhibits tyrosine kinase activity of Bcr-Abl but cannot give a free remission of the disease. We investigated the role of the $Ca^{2+}$ signaling in Bcr-Abl-expressing cells in order to find new pharmacological targets in LSC.

In this study, we described the different $Ca^{2+}$ influxes in 32d-p210 cells compared to wild type (WT) cells: $Ca^{2+}$ entries in response to thrombin or to pharmacological store depletion are reduced in Bcr-Abl-expressing cells. The decrease of SOCE is not related to Orai1/STIM1 proteins expression but most probably to a change of stoichiometry. In Bcr-Abl-expressing cells, the decrease of SOCE is also accompanied by a reduction of the Nuclear Factor of Activated T cells (NFAT) translocation and regulates cell proliferation and migration. Tyrosine kinase activity of Bcr-Abl does not regulate directly $Ca^{2+}$ homeostasis but participates to the general disorganization of cell function in leukemia cells notably via Protein Kinase C (PKC).

## RESULTS

### Calcium influxes in Bcr-Abl-expressing cells

In order to study the Bcr-Abl-dependent $Ca^{2+}$ homeostasis, we characterized the different $Ca^{2+}$ influxes in 32d cells wild type (32dWT) or stably transfected with the

p210$^{bcr-abl}$ oncogene (32d-p210). For all the Ca$^{2+}$ experiments, 32d cells (WT and -p210) were immobilized on fibronectin-coated coverslips and the ratiometric Ca$^{2+}$ indicator dye Fura-2-Acetoxymethyl ester (Fura-2 AM) was used to analyze Ca$^{2+}$ variation of single cells. We first studied the basal Ca$^{2+}$ "leak". To measure a constitutive Ca$^{2+}$ influx, cells were incubated very briefly (30 or 40 seconds) in an extracellular 0 mM Ca$^{2+}$ solution and quickly changed to 1.8 mM Ca$^{2+}$ buffer (Figure 1A). With this protocol, a weak decrease of the ratio of fluorescence during the incubation of 0 mM Ca$^{2+}$ buffer was observed, showing the basal Ca$^{2+}$ entry in resting cells. Then, the 1.8 mM Ca$^{2+}$ buffer incubation allowed the return to the basal level, suggesting that no other Ca$^{2+}$ channels were activated after this phase (Figure 1A). To increase the gradient toward the membrane, the same experiments were performed with a 5 mM Ca$^{2+}$ instead of 1.8 mM Ca$^{2+}$. In our cells lines, the constitutive Ca$^{2+}$ influx was weak in presence of 1.8 or 5 mM Ca$^{2+}$ buffer and may not play a predominant role in Ca$^{2+}$ homeostasis. Moreover, no difference was measured between WT and Bcr-Abl-expressing cells (Figure 1B) suggesting that no constitutive Ca$^{2+}$ entry is increased contrary to what has been observed in other types of cancer cells [33].

To investigate the GPCR activated pathways, cells were treated with 1 U/ml of thrombin. The thrombin-evoked intracellular Ca$^{2+}$ responses were characterized by a quick peak with a rapid rising phase, followed by a longer sustained phase during which the intracellular Ca$^{2+}$ remained relatively high and slowly decreased (Figure 1C). The rapid phase of the Ca$^{2+}$ increase was analyzed by the maximum of the peak (maximum of response) while the half-time of response was used for evaluating the duration of the sustained phase (Figure 1D). The maximum peak decreased slightly in 32d-p210 cells compared to 32dWT and the sustained phase showed a strong reduction in 32-p210 cells (Figure 1C and 1D). In conclusion, Bcr-Abl expression induced a decrease of thrombin-dependent Ca$^{2+}$ response. This experiment measured a global cytosolic Ca$^{2+}$ signal but could not distinguish between intracellular stock release and entry through PM. To understand the thrombin-dependent Ca$^{2+}$ response, cytosolic Ca$^{2+}$ variations were observed in presence of extracellular 0 mM Ca$^{2+}$ solution (Figure 2A). In these conditions, the peak of response increased in 32dWT showing that this first phase depends mainly on Ca$^{2+}$ intracellular stock release and not on an extracellular entry (Figure 2B). Moreover, the small gain could be due to a reaction of the cells stressed by a lack of extracellular Ca$^{2+}$. However, the incubation of 32d-p210 cells in 0 mM Ca$^{2+}$ leaded to a small decrease of the maximum, which turned back to the 32dWT level in 1.8 mM Ca$^{2+}$ buffer (Figure 2B). Besides, the half-time of response was strongly decreased in both cells lines in 0 mM Ca$^{2+}$ buffer. This result shows that the second phase of the intracellular response mainly involves the extracellular Ca$^{2+}$ entries (Figure 2C). The reduction of the slow sustained phase was more pronounced in control cells compared to Bcr-Abl-expressing cells. This suggests that the longer sustained phase in 32dWT cells is due to higher level of Ca$^{2+}$ entries. Because the main Ca$^{2+}$ influx in non-excitable cells is SOCE, their implication in the sustained phase was investigated. We did not succeed in efficiently transfecting both cell lines so we used two pharmacological Ca$^{2+}$ influxes inhibitors: YM-58483 and SKF-96365 (Figure 2D). In 32dWT, treatment with inhibitors resulted in a strong decrease of the half-time of thrombin-dependent Ca$^{2+}$ response (Figure 2D and 2F) reaching the value to that of the 32d-p210 cells in normal conditions. YM-58483 and SKF-96365 induced a slight reduction of the half-time of response in 32d-p210. This result shows that the difference in half-time of thrombin response between control and Bcr-Abl-expressing cells is due to Ca$^{2+}$ entry through PM Ca$^{2+}$ channels. This suggests that a reduction of SOCE in Bcr-Abl-expressing cells is responsible of the decreased sustained phase during the thrombin-elicited Ca$^{2+}$ signal. The maximum of response did not show notable changes in WT cells during inhibitors incubation in contrast to the Bcr-Abl-expressing cells (Figure 2D and 2E). Same results were obtained using another SOC inhibitor: GSK-7975A (Supplementary Figure 1C). In these latter cells, YM-58483 or SKF-96365 treatment induced a decrease of the maximum of response suggesting that the increase of the peak measured in 32d-p210 cells (Figure 1D) could be due to Ca$^{2+}$ entry through PM.

**SOCE is reduced in Bcr-Abl-expressing cells**

Because thrombin signaling induces several Ca$^{2+}$ pathways, we investigated SOCE with specific protocols. Incubation with a sarco/endoplasmic reticulum Ca$^{2+}$-ATPase (SERCA) inhibitor (cyclopiazonic acid (CPA)) induces the decrease of ER Ca$^{2+}$ concentration and then the activation of SOCs (Orai1 and TRPC1) via STIM1 dimerization [14, 20]. The ER Ca$^{2+}$ release was analyzed by the maximum and the quantity (area under the curve) of cytoplasmic Ca$^{2+}$ increase in CPA-0 mM Ca$^{2+}$ conditions (Figure 3A and 3B). With this protocol, the ER depletion maximum and amplitude were lower in Bcr-Abl-expressing cells compared to control cells, as previously described and linked to a B-cell lymphoma 2 protein (Bcl-2) independent inhibition of apoptotic signaling [31]. After ER depletion induced SOCs activation, incubation with 1.8 mM Ca$^{2+}$ solution was followed by a Ca$^{2+}$ entry through SOCs. This Ca$^{2+}$ influx was characterized by the rate of entry (measured by the slope of the curve) and the maximum of cytosolic Ca$^{2+}$ increase (maximum). Bcr-Abl-expressing cells displayed a decrease in the rising slope and in the maximum of Ca$^{2+}$ entry, compared to 32dWT cells (Figure 3C). Same results were obtained after the incubation with ER Ca$^{2+}$ chelator (N,N,N',N'-tetrakis(2-pyridylmethyl)ethane-1,2-diamine (TPEN)) that mimicked the decrease of ER Ca$^{2+}$ concentration (Figure 3C). SOCEs in 32d cells were also confirmed by the incubation of the inhibitor GSK-7975A (Supplementary Figure 1A, 1B) These results show that Bcr-Abl expressing cells display reduced SOCE.

To understand the origin of SOCE decrease in Bcr-Abl-expressing cells, the expression of the major SOC components was studied and compared to control cells. We could not detect any difference in the Ribonucleic Acid (RNA) levels of STIM1 and Orai1 between 32dWT and 32d-p210 cells (data not shown). Western Blot analyses revealed no difference of expression for the ER $Ca^{2+}$ sensor STIM1 whereas an increase of Orai1 expression was shown in 32d-p210 cells (Figure 4A and 4B). Furthermore, TRPC1 expression was decreased in these cells (Figure 4A and 4B). The decrease of TRPC1 expression is consistent with the reduced SOCE in 32d-p210 cells (Figure 3). Moreover, the STIM1 amounts seem not to be dependent on the Bcr-Abl expression in contrast to Orai1. The increase of the Orai1 expression probably leads to a change of the stoichiometry between STIM1 and Orai1, that could also explain the reduced SOCE in 32d-p210 cells [34, 35].

Bcr-Abl is a constitutively active tyrosine kinase and regulation of STIM1, Orai1 and TRPC1 by phosphorylation has been well documented. For instance, PKC-induced phosphorylation of TRPC1 is known to enhance SOCE [36, 37], and Orai1 is also known to be inhibited by serine/threonine phosphorylation [38]. In order to study phosphorylation states of these proteins, Phos-Tag™ molecules were introduced into SDS-PAGE gel. This molecule binds the phosphate moiety of proteins and the in-gel complex formed is retained during sample electrophoresis. Each Phos-Tag™ molecule binds one phosphate and in consequence, each phosphorylation form of the protein can be isolated and visualized by Western Blot [39]. The use of Phos-Tag™ molecule modifies the

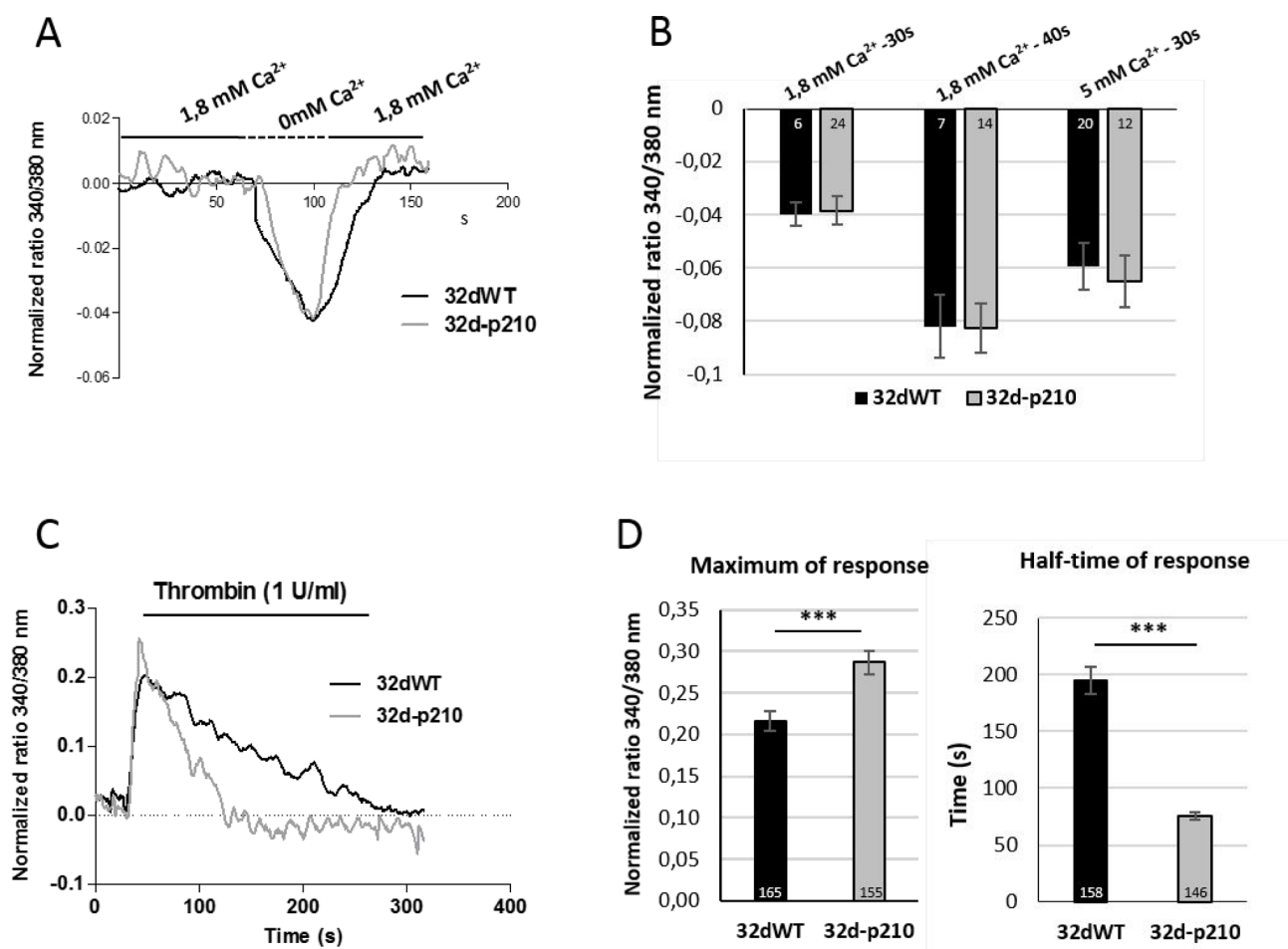

**Figure 1: Calcium entries in 32d cells.** (**A**) Constitutive entries in 32dWT (dark) and 32d-p210 (grey) cells. Cells were plated on fibronectin-coated coverslips and loaded with Fura-2 AM. After incubation in 1.8 mM $Ca^{2+}$ buffer, cells were perfused with 0 mM $Ca^{2+}$ buffer for 30 seconds. Cytosolic $Ca^{2+}$ variations were recorded by ratiometric fluorescence at 340/380 nm. (**B**) Quantification of constitutive $Ca^{2+}$ entries in 32dWT and 32d-p210 cells. Cells were incubated in 1.8 or 5 mM $Ca^{2+}$ solution and perfused with 0 mM $Ca^{2+}$ buffer for 30 or 40 seconds. The 340/380 nm ratio between the peak of decrease and the basal value has been measured. (**C**) Thrombin-induced $Ca^{2+}$ entry in 32dWT (black line) and 32d-p210 (grey line) cells. Cells were plated on fibronectin-coated coverslips and incubated with 1 U/ml thrombin in 1.8 mM $Ca^{2+}$ buffer. (**D**) Quantification of thrombin-induced $Ca^{2+}$ entry in 32dWT (dark) and 32d-p210 (grey) by measured the peak of response (maximum in 340/380 nm fluorescence ratio) and the half time of response (in seconds). Bar graphs represent mean rates response ± SEM. ***$P < 0.001$.

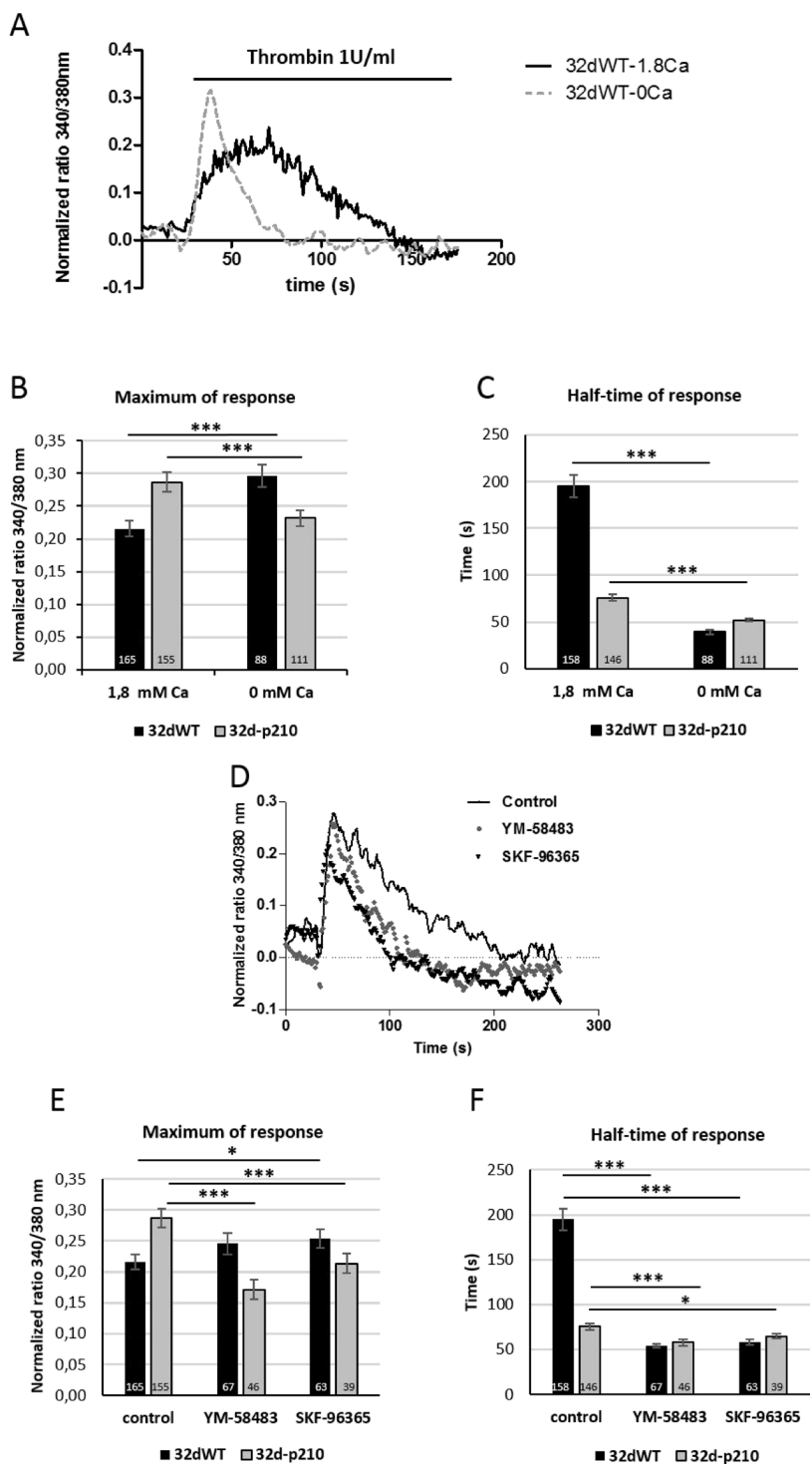

**Figure 2: Characterization of thrombin-induced calcium entries.** (**A**) Thrombin-induced $Ca^{2+}$ entry in 32dWT cells in 1.8 mM $Ca^{2+}$ buffer (solid line) or 0 mM $Ca^{2+}$ buffer (dotted line) in presence of 1 U/ml thrombin. Cells were plated on fibronectin-coated coverslips and loaded with Fura-2 AM. Cytosolic $Ca^{2+}$ variations were recorded by ratiometric fluorescence at 340/380 nm. (**B** and **C**) Quantification of thrombin-induced $Ca^{2+}$ entry in 1.8 mM $Ca^{2+}$ buffer or 0 mM $Ca^{2+}$ buffer in presence of 1 U/ml thrombin. The peak of response (maximum in 340/380 nm fluorescence ratio) and the half-time of response (in seconds) were measured in 32dWT (dark) and 32d-p210 (grey) cells. The control values are pooled from different experiments. (**D**) Thrombin-induced $Ca^{2+}$ entry in 32dWT cells after treatment with $Ca^{2+}$ channels blockers (YM-58483 and SKF-96365). Cells were pre-incubating or not (solid line) with 10 μM YM-58483 (circle) or 40 μM SKF-96365 (triangle) during 5 minutes and incubated with 1 U/ml thrombin in 1.8 mM $Ca^{2+}$ buffer. (**E** and **F**) Quantification of thrombin-induced $Ca^{2+}$ entry with or without pre-incubation with 10 μM YM-58483 or 40 μM SKF-96365 during 5 minutes and incubated with 1 U/ml thrombin in 1.8 mM $Ca^{2+}$ buffer. Measurements are represented as the peak of response (maximum in 340/380 nm fluorescence ratio) and the half-time of response (in seconds). Bar graphs represent mean rates ± SEM. ***$P < 0.001$.

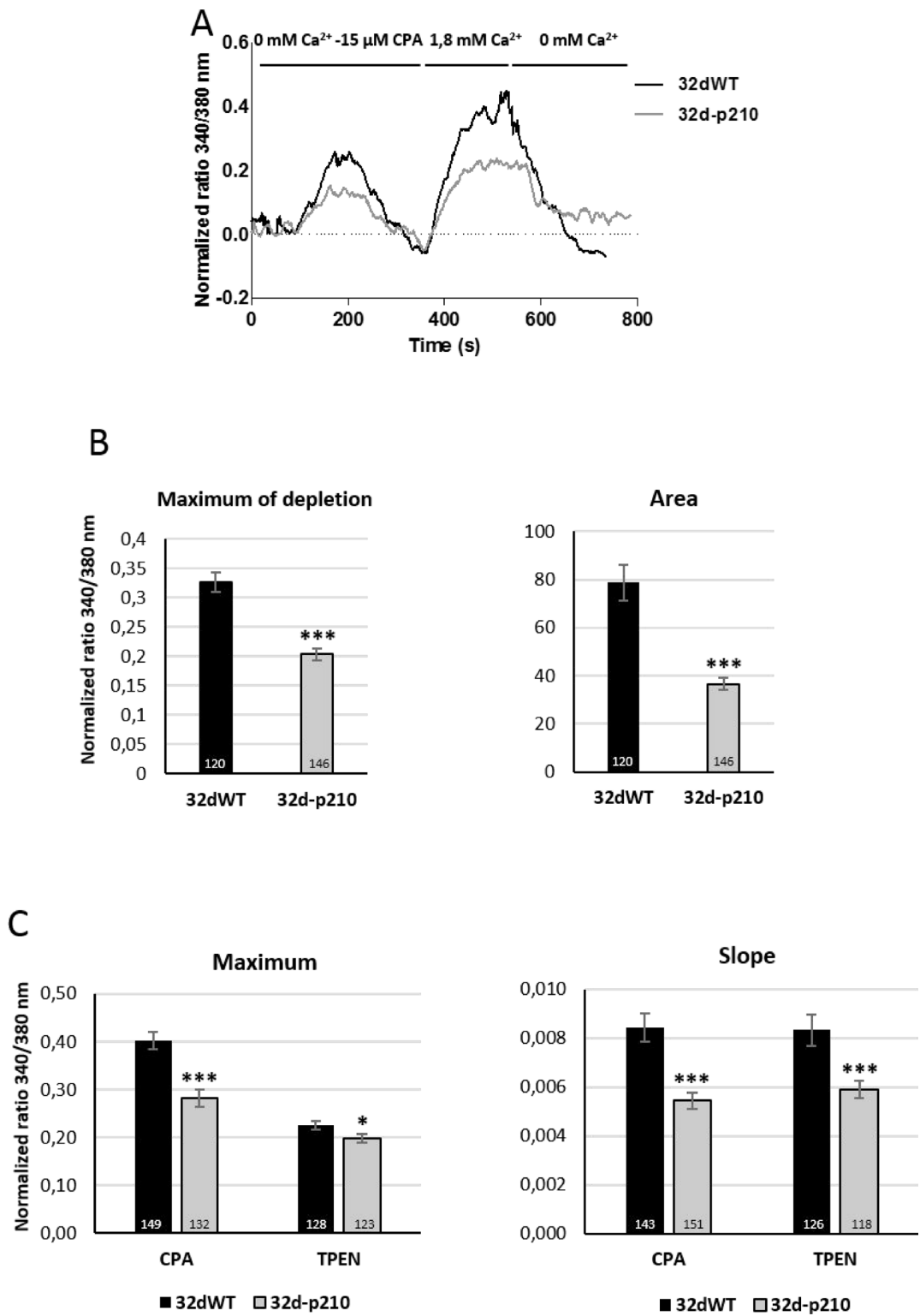

**Figure 3: Store-Operated Calcium Entries (SOCE).** (**A**) Store-operated $Ca^{2+}$ entry in 32d cell lines in 32dWT (dark line) and 32d-p210 (grey line). The ER depletion were trigged by 15 μM CPA (SERCA inhibitor) in 0 mM $Ca^{2+}$ solution which allowed SOCE in presence of 1.8 mM $Ca^2$ prior to a 0 mM $Ca^{2+}$ buffer incubation. (**B**) Quantification of ER depletion with the maximum of depletion (340/380 nm fluorescence ratio) and the amount of $Ca^{2+}$ release (area under the curve) after normalization of the response. (**C**) SOCE measurements in 32dWT (dark) and 32d-p210 (grey) cells with the maximum (340/380 nm fluorescence ratio; left panel) and the speed of entries (initial slope; right panel). SOCE were induced by ER depletion with the 0 mM $Ca^{2+}$-15 μM SERCA inhibitor CPA or with 0 mM $Ca^{2+}$-1 mM ER $Ca^{2+}$ chelator TPEN for 5 minutes. Bar graphs represent mean rates ± SEM. ***$P < 0.001$.

www.oncotarget.com                                                                 26314                                                                     Oncotarget

sample migration into the gel and then apparent molecular weight is not reliable. These experiments allow a direct observation of phosphoproteins as well as their levels of phosphorylation. Therefore, 32dWT and 32d-p210 lysates were treated or not with CPA in order to activate SOCs. The dephosphorylated form of the protein was used as a control (Figure 4C). In 32dWT and 32d-p210 cells, TRPC1 presented a dephosphorylated form but no difference in normal or stimulated conditions could be observed. The tyrosine phosphorylation of STIM1 is required for Orai1 interaction in ER/PM junction (for review: [40, 41]). As expected, the CPA stimulation induced a STIM1 phosphorylation in both 32d cell lines, and we observed the presence of several phosphorylated forms of Orai1. Nevertheless, no statistical variations were measured in the ratio phosphorylated/non-phosphorylated Orai1 protein as well as in STIM1 and TRPC1 that could explain the reduced SOCE measured in Bcr-Abl-expressing cells. These data suggest that no direct or indirect Bcr-Abl-dependent phosphorylation of main SOC components is involved in the observed SOCE decrease.

### The role of SOCE in 32d cells

In order to understand the functional consequences of SOCE decrease in Bcr-Abl-expressing cells, several $Ca^{2+}$-regulated processes involved in leukemia development were studied. Nuclear factor of activated T cells (NFAT) is a $Ca^{2+}$-dependent transcriptional factor that regulates various aspects of cellular function (including migration and proliferation). A cytoplasmic $Ca^{2+}$ increase (via SOCs or ROCs) induces activation of the $Ca^{2+}$-dependent phosphatase, calcineurin which dephosphorylates NFAT causing its translocation into the nucleus (for review: [42]). A long lasting $Ca^{2+}$ response can maintain NFAT into the nucleus, which could be the support of differences in $Ca^{2+}$-dependent transcription between 32dWT and 32d-p210 cells that display shorter $Ca^{2+}$ responses to thrombin. The NFAT activation and translocation can be quantified by analyzing nucleus/cytoplasm ratio of NFAT means intensity. Cells were treated with or without CPA incubation in presence of calcineurin inhibitor cyclosporine A (CsA) or SOCE inhibitors, fixed and labelled with anti-NFAT2 antibodies (Figure 5A). In control conditions 32dWT cells, 50% of NFAT2 signal was in the nucleus. In contrast in 32d-p210 cells, NFAT2 signal was stronger in the cytoplasm (Figure 5B). After SOCs activation by CPA, NFAT2 translocated into the nucleus in both cell lines although more protein translocated into the nucleus of 32dWT cells compared to 32d-p210, a result consistent with the differences already observed in SOCE (CPA protocol) and ROCE (thrombin protocol) between the two cell lines (Figures 1 and 3). The use of SOCE inhibitors reduced NFAT2 translocation, confirming the link between SOCE and NFAT activation. Same results were obtained with NFAT1 (data not shown).

Altogether, decreased $Ca^{2+}$ influx in stimulated Bcr-Abl-expressing cells led to a reduction of the activation of the NFAT $Ca^{2+}$-dependent pathway.

We next explored the impact of SOCE on proliferation of 32d cells. Bcr-Abl induces an increase of proliferation in leukemia cells mainly through its tyrosine kinase activity as already described [43] (Figure 5C). As expected, SOCE inhibition reduced also the rate of proliferation in 32d cells at 48 and 72 hours (Figure 5C). In fact, SOCE is known to regulate proliferation (for review: [27]) via several pathways (Extracellular signal-Regulated Kinases (ERK), NFAT, etc.) and to modify transcription of genes involved in cell cycle. We showed that cell proliferation could also be reduced by treatments with $Ca^{2+}$ channels blockers in 32d-p210 cells, suggesting that these inhibitors can counteract the positive effect of Bcr-Abl on cell cycle. These results (Figure 5C) suggest that $Ca^{2+}$ entries could play a role in cells proliferation but the main regulator remains the constitutive tyrosine kinase activity of Bcr-Abl.

Since Bcr-Abl expression has also been involved in activation of amoeboid cell migration [44, 45], the role of $Ca^{2+}$ homeostasis in migration was also studied. 2D migration experiments were carried in fibronectin layer and length (distance traveled) and speed were measured (Figure 5D). Both factors were higher in 32d-p210 cells (Figure 5D), which confirms the enhanced spontaneous migration of Bcr-Abl-expressing cells as previously described [44]. The pharmacological inhibition of SOCE induced an increase of spontaneous cell migration in 32-p210 cells (length and speed). SOCE has been reported to regulate migration in several previous studies, but the inhibition of SOCE usually reduced motility of cancer cells (for review: [27, 46]). Nevertheless several recent studies described that decrease of cytosolic $Ca^{2+}$ could lead to an increase of cell migration [47, 48]. Similarly, the inhibition of $Ca^{2+}$ channels in our model of hematopoietic progenitors induced an increase of cell motility. This suggests that the reduction of SOCE in Bcr-Abl-expressing cells is participating (together with RhoA activation for instance) to the enhanced spontaneous migration of these progenitors.

### Imatinib and calcium influxes

Imatinib is the first line TKI used to treat CML patients. We used imatinib to study the link between $Ca^{2+}$ influx and kinase activity of Bcr-Abl. In thrombin-stimulated 32-p210 cells, imatinib induced only a small decrease of the half-time of response compared to non-treated cells (Figure 6A and 6B). In contrast, imatinib incubation strongly increased SOCE (maximum and slope) in 32-p210 cells to a level comparable with the one measured in 32dWT cells (Figure 6C and 6E). Moreover, TKI did not change the ER depletion (maximum and area) (Figure 6D) in 32d-p210 but restored the SOCE. These results suggest that Bcr-Abl inhibits SOCE through TK activity directly or indirectly (Figure 6E). The inhibition of SOCs by YM-58483 reduced SOCE as expected and only a weak additional effect could be measured in the presence of imatinib suggesting that tyrosine kinase activity regulates directly SOCE.

We next tried to find a link between the Bcr-Abl expression and the deregulation of $Ca^{2+}$ homeostasis. PKC proteins could represent a good candidate because they have been previously described to be deregulated in Bcr-Abl-expressing cells [49] and PKC can modulate SOCE in various cells [48, 50]. Inhibition of PKC with the non-selective inhibitor Bisindolylmaleimide (BIM) induced a decrease of SOCE (slope and maximum) in 32dWT cells but no significant effect could be measured in 32d-p210 cells (Figure 7A). $Ca^{2+}$ entries in 32dWT cells seem to be dependent on PKC activity in contrast to Bcr-Abl-expressing cells. This suggests that the PKC-dependent pathway activating SOCE in control cells is already inhibited by Bcr-Abl in leukemia cells. Moreover, thrombin-stimulated $Ca^{2+}$ entry was also dependent on PKC activity in 32dWT cells. In these conditions, BIM treatment induced a strong decrease of half-time of response comparable to the one observed when measuring only SOCE (Figure 7B). No strong difference was observed in half-time of response in 32d-p210 cells after BIM treatment. This suggests that in 32dWT cells, PKC could principally enhance SOCE, and in turn the duration of intracellular $Ca^{2+}$ signal in response to thrombin.

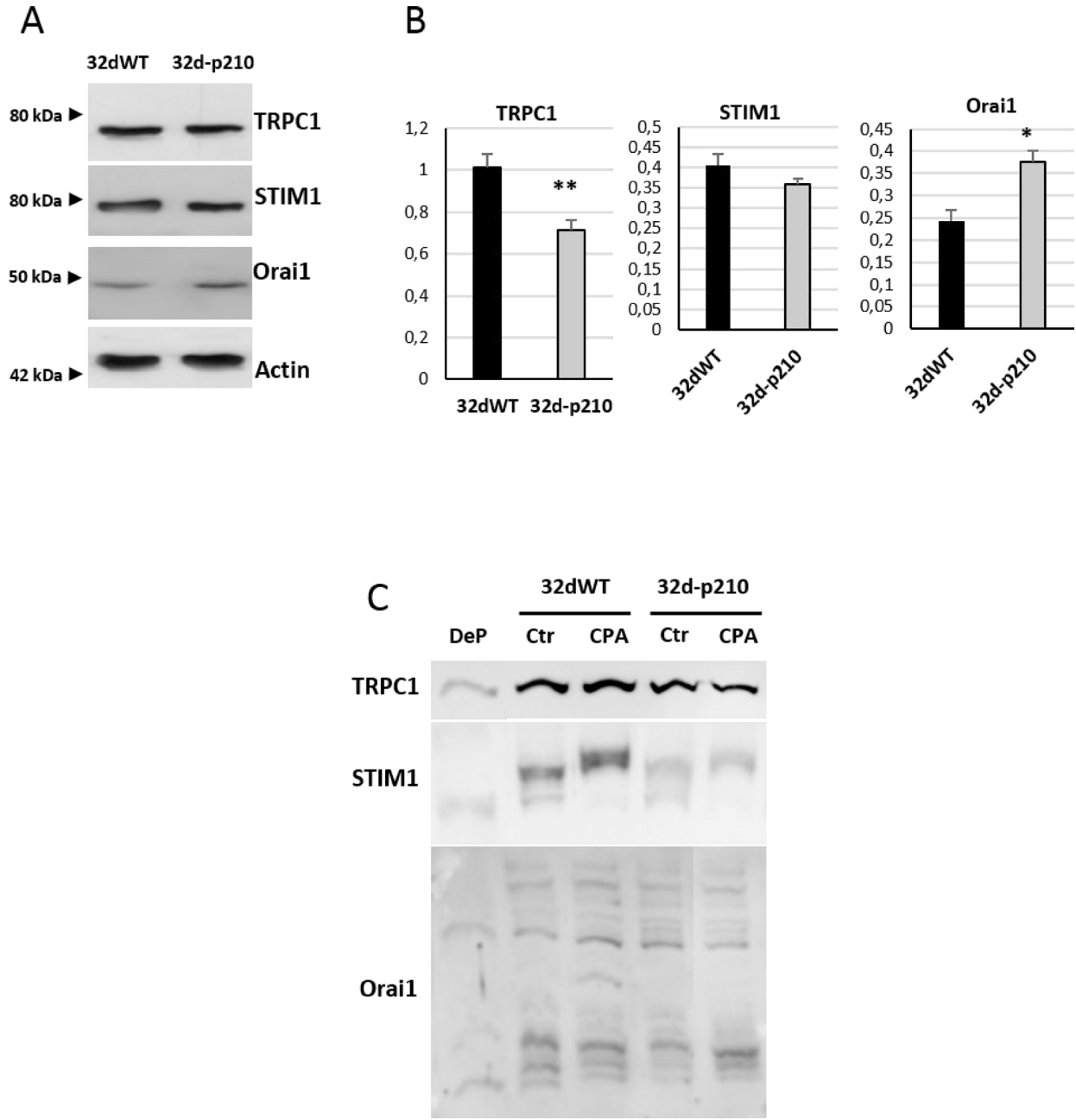

**Figure 4: SOC constituents in 32d cells.** (**A**) Expression of TRPC1, Orai1 and STIM1 in 32dWT and 32d-p210 cells. Cell lysates were loaded in 8% SDS-PAGE gel and western blot analysis were performed using β-actin as a loading control. (**B**) Quantification of proteins expression from 13 independent experiments using GeneTools image analysis 4.3.5 Software. $^{**}P < 0.005$. (**C**) Phosphorylation state of TRPC1/Orai1/ STIM1. Phosphorylated proteins were revealed by the PhosTag™ molecules, which are able to interact with phosphate. PhosTag™ were introduced to the separation SDS-PAGE gel and delayed phosphoprotein migration. Each phosphorylated form could be separated and visualized with specific antibodies in Western Blot. Dephosphorylated (DeP) form of proteins of interest was used as a negative control. Cells were lysate after incubation in 1.8 mM $Ca^{2+}$ buffer (Ctr) or after ER depletion (0 mM $Ca^{2+}$-15 μM CPA during 7 minutes) (CPA).

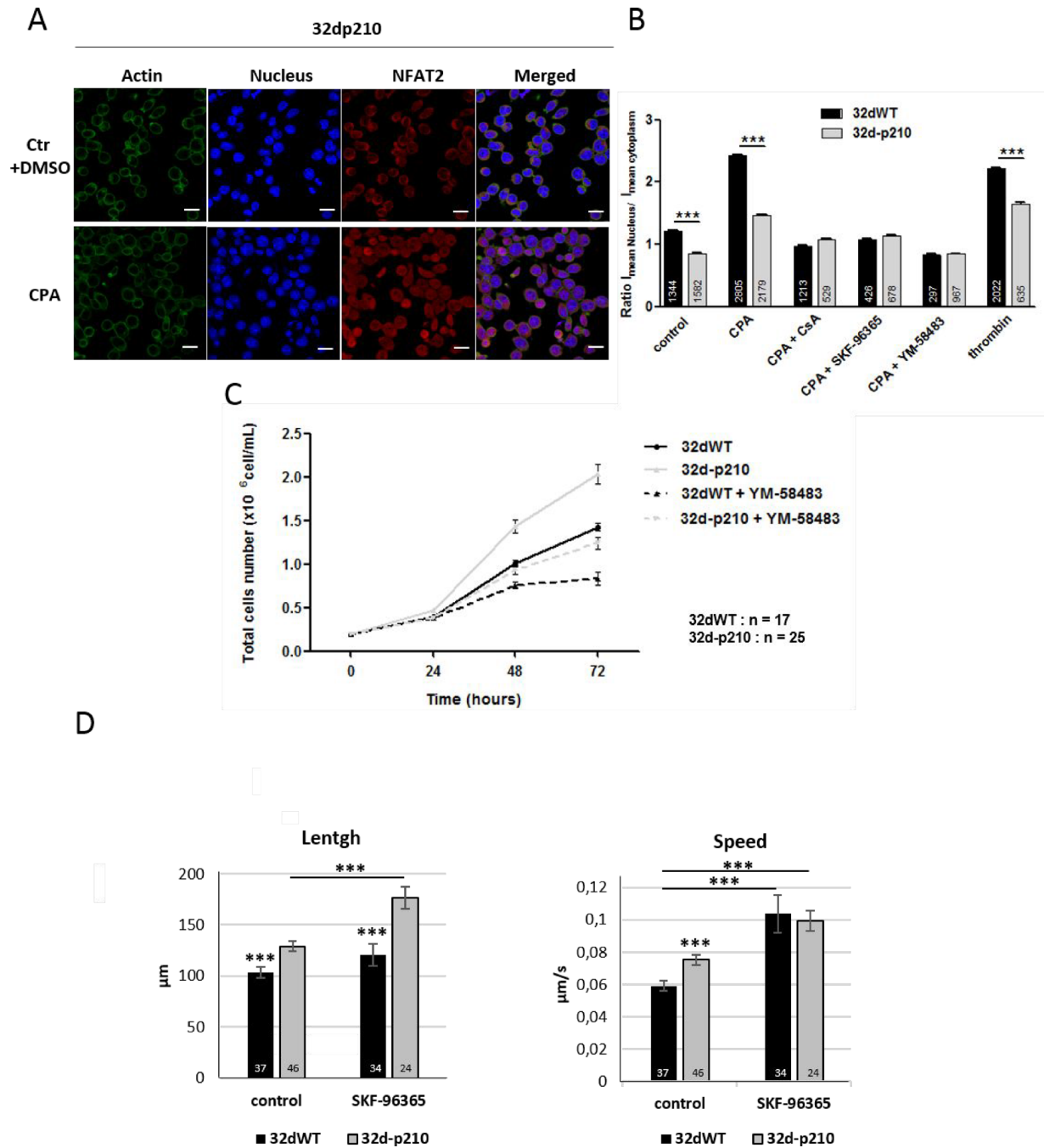

**Figure 5: SOCE dependent cellular processes.** (**A**) NFAT2 translocation in 32d-p210 cells. Immunostaining of actin (Phalloidin, green), nucleus (TOPRO, blue) and NFTA2 (red) were done in 1.8 mM $Ca^{2+}$ buffer + DMSO (ctr) or in 1.8 mM $Ca^{2+}$ buffer-15 μM CPA (CPA) for 30 minutes". Images were obtained using confocal microscopy; bar = 10 μm. (**B**) Quantification of NFAT2 translocation from cytoplasm to nucleus by analyzing nucleus/cytoplasm ratio of NFAT2 means intensity in 32dWT (dark) and 32d-p210 (grey) cells. Cells were fixed after incubation in 1.8 mM $Ca^{2+}$ buffer + DMSO (ctr) for 30 minutes or after stimulation with 0 mM $Ca^{2+}$ buffer-15 μM CPA for 7 minutes prior to 1.8 mM $Ca^{2+}$ buffer-15 μM CPA for 30 min (CPA). Alternatively, cell were stimulated with 1 U/ml thrombin for 30 minutes. Cells were pre-treated with 1 μM CsA for 30 minutes (CPA+CsA), or treated with 40 μM SKF-96365 (CPA+SKF-96365) or 10 μM YM-58483 (CPA+YM-58483) during incubation 1.8 mM $Ca^{2+}$-15 μM. CPA for 30 minutes. Bar graphs represent mean rates of nucleus/cytoplasm ratio of NFAT2 means intensity ± SEM. ***$P < 0.001$. (**C**) Proliferation rate of 32dWT and 32d-p210 in normal medium with or without 10 μM YM-58483. Cells viability was estimated with trypan blue coloration at 24, 48 and 72 hours of incubation. (**D**) Cell migration was quantified by tracking analyzed using Imaris software. Total length and mean speed of each cell were measured in normal medium supplemented or not with 40 μM SKF-96365. Cell migration was recorded at 37° C for 30 minutes using the spinning disk confocal microscope (Olympus IX81-ZDC).

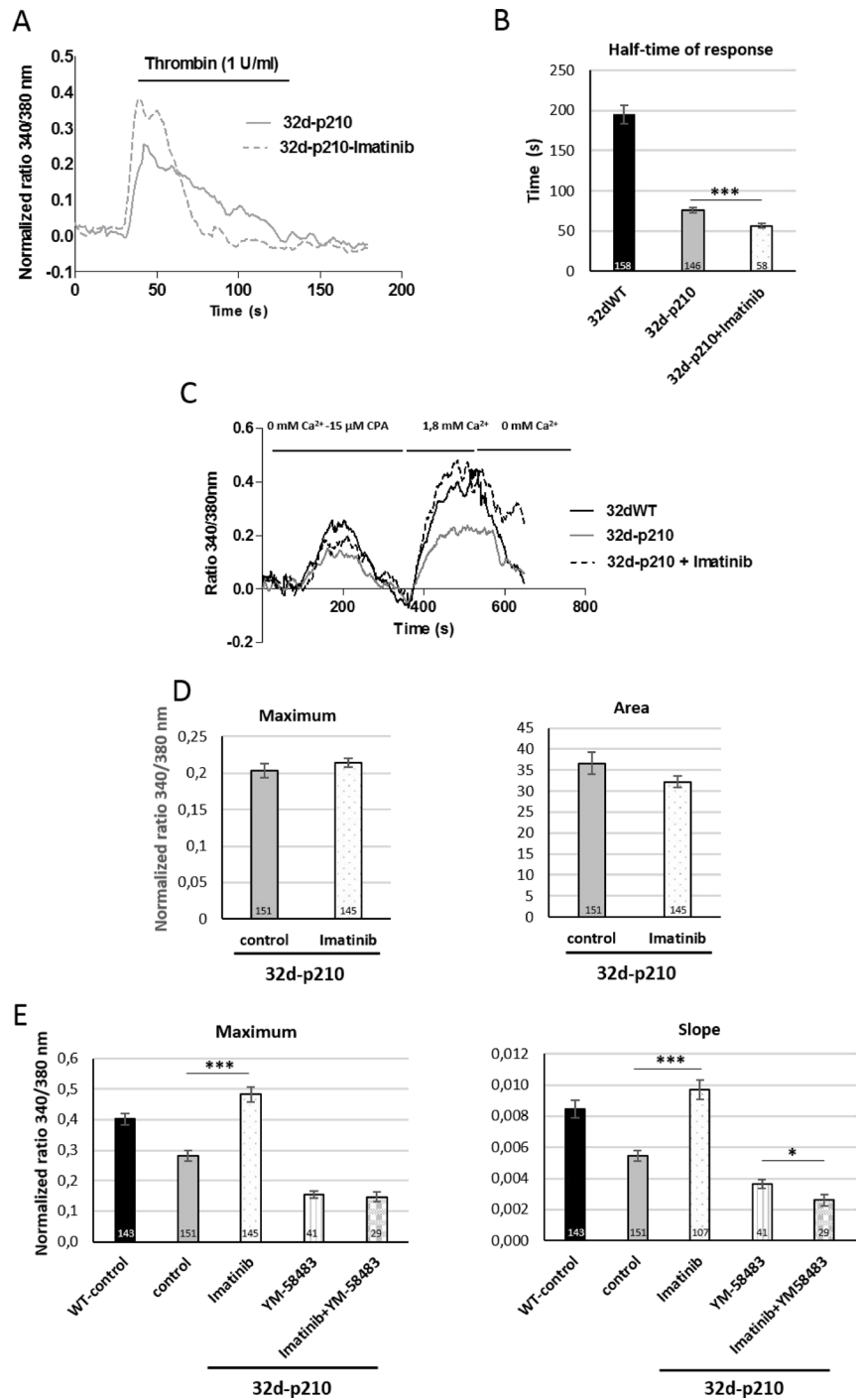

**Figure 6: Role of Bcr-Abl tyrosine kinase activity.** (**A**) Thrombin-induced $Ca^{2+}$ entry in imatinib pretreated 32d-p210 cells. Cells were plated on fibronectin-coated coverslips and treated (dotted line) or not (solid line) with 10 μM imatinib for 4 hours and next incubated in 1.8 mM $Ca^{2+}$ buffer-1 U/ml thrombin. (**B**) Quantification of half-time of response of thrombin-induced $Ca^{2+}$ entry in imatinib pretreated 32d-p210 cells. Bar graphs represent mean rates of half-time of response ± SEM in 32dWT (dark), in non-treated 32d-p210 (grey) and treated 32d-p210 (white) cells. The control values are pooled from different experiments. ***$P < 0.001$. (**C**) SOCE in imatinib pretreated 32d-p210 cells. 32d-p210 cells were treated (dotted line) or not (grey line) with 10 μM imatinib for 4 hours. ER depletion was trigged by 15 μM CPA in 0 mM $Ca^{2+}$ solution and SOCE were visualized with 1.8 mM $Ca^{2+}$ buffer incubation. Same protocol was performed to 32dWT cells (dark) as control. (**D**) Quantification of ER depletion in non-treated 32d-p210 (grey) and treated 32d-p210 (white) cells with the maximum of depletion (340/380 nm fluorescence ratio, left panel) and the amount of $Ca^{2+}$ release (area under the curve, right panel) after normalization of the response. (**E**) SOCE quantification in 32d cells treated or not with imatinib and/or $Ca^{2+}$ channels blockers. 32d-p210 cells were treated or not with 10 μM imatinib for 4 hours and 10 μM YM-58483 were injected or not before ER depletion. The maximum (340/380 nm fluorescence ratio; left panel) and the speed of entries (initial slope; right panel) were measured. Bar graphs represent mean rates ± SEM. ***$P < 0.001$.

## DISCUSSION

In this study, we explored the deregulation of $Ca^{2+}$ homeostasis, particularly SOCE, in hematopoietic cells expressing $p210^{bcr-abl}$ oncogene. We showed that in 32d-p210 cells, thrombin-dependent and store-dependent $Ca^{2+}$ entries were significantly reduced compared to control cells. Interestingly, the participation of constitutive $Ca^{2+}$ entry to the regulation of $Ca^{2+}$ homeostasis was negligible, in contrast to breast cancer cells [33]. In agreement with previous works [31, 51], ER $Ca^{2+}$ store depletion was also reduced in 32d-p210 cells (Figure 3B). Thrombin-induced entries were characterized by two phases: a first peak followed by a sustained phase. This work showed that, following thrombin-elicited ER $Ca^{2+}$ release, SOCE participated to the second sustained phase in both cell lines. However, this second phase was significantly shorter in Bcr-Abl-expressing cells, which is partly due to a reduction of SOCE and may contribute to the alteration of downstream $Ca^{2+}$-dependent signaling pathways. Particularly, it is well known that a sustained $Ca^{2+}$ response supported by SOCE is more efficient in promoting NFAT activation and translocation, than a transient $Ca^{2+}$ increase. Surprisingly the first phase of thrombin-induced response (peak), mainly due to $Ca^{2+}$ release, was also slightly dependent on extracellular $Ca^{2+}$ only in Bcr-Abl-expressing cells. This suggests that $Ca^{2+}$ entry may also participate to the resulting peak value due to changes in $Ca^{2+}$ entry pathways.

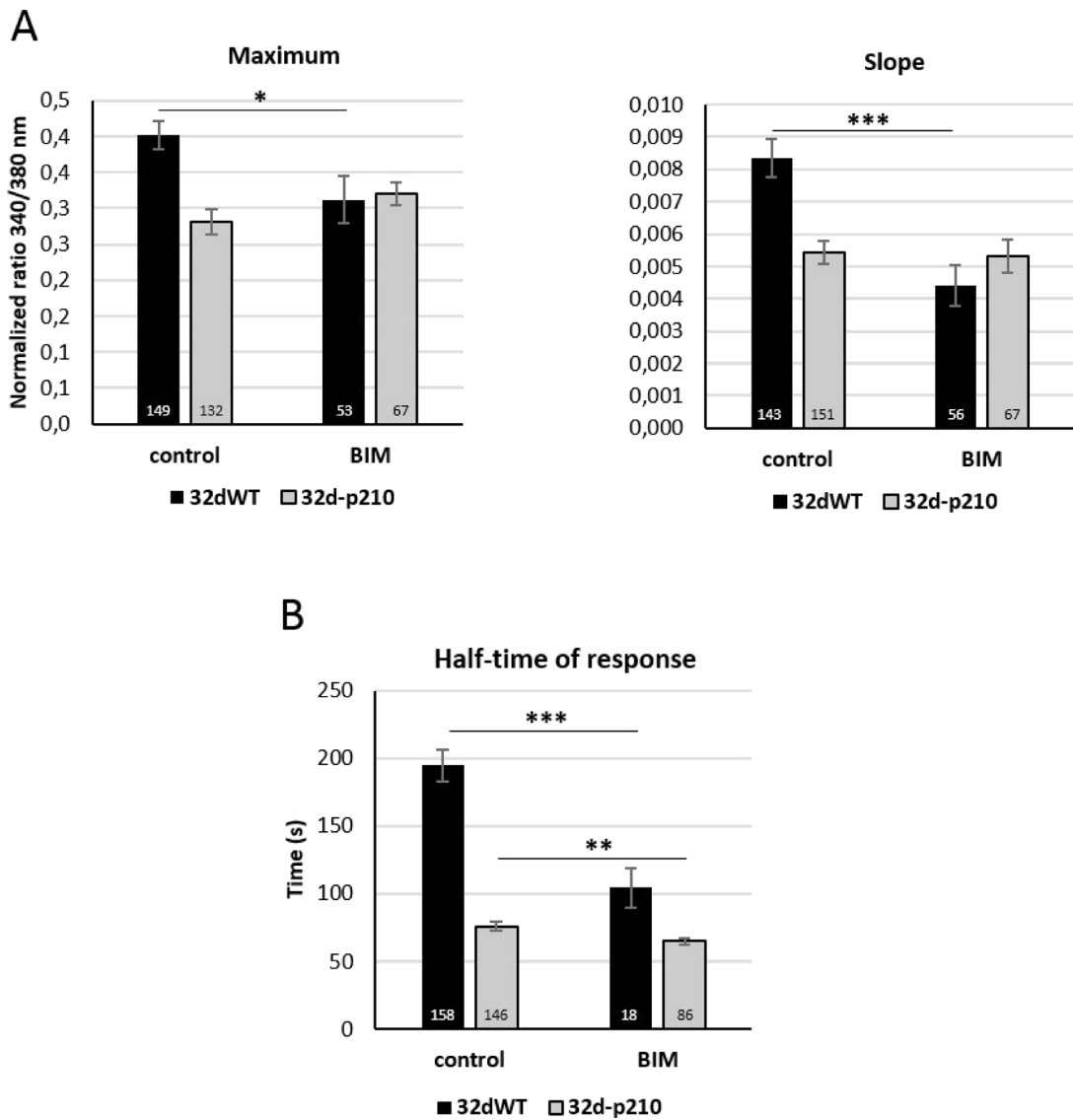

**Figure 7: Role of PKC in calcium influxes.** (**A**) SOCE quantification with or without PKC inhibitor (BIM) in 32dWT (dark) and 32d-p210 (grey) cells. Cells were incubated or not with a PKC inhibitor BIM (10 μM for 30 minutes before CPA treatment) in 1.8 mM $Ca^{2+}$ buffer. SOCs activation protocol was performed and SOCE were measured by the maximum (340/380 nm fluorescence ratio; left panel) and the speed of entries (initial slope; right panel). (**B**) Thrombin-induced $Ca^{2+}$ entry quantification with or without PKC inhibitor (BIM) in 32dWT (dark) and 32d-p210 (grey) cells (half-time of response in seconds). Cells were incubated or not in 1.8 mM $Ca^{2+}$ -10 μM BIM for 30 minutes before CPA treatment, and 1.8 mM $Ca^{2+}$-1 U/ml thrombin buffer was added. Bar graphs represent mean rates ± SEM. ***$P < 0.001$.

To understand the mechanism of SOCE reduction in leukemia cells, we explored expression levels of SOC constituents in comparison to control cells. No difference in STIM1 expression could be evidenced, but Orai1 expression was enhanced in 32d-p210 cells while TRPC1 expression was diminished. The molecular interaction between STIM1 and Orai1 is now well-documented [52]: Orai1 was initially described to operate as a tetramer channel [11] and recently demonstrated to function as a hexamer [53]. Increase in Orai1 expression leads to a decrease in CRAC channel activity by reducing STIM1/Orai1 binding stoichiometry [34, 35]. In addition, after its activation, Orai1 is inactivated in a fast $Ca^{2+}$-dependent manner. This rapid $Ca^{2+}$-dependent inactivation is dependent on Orai1 expression level. Indeed, in our cell model, the increase of Orai1 expression combined with a stable STIM1 expression could explain the decrease of SOCE in 32d-p210 (Figures 2 and 3). Thus, Orai1 stoichiometry is crucial for its different modes of action. We also observed a decrease in TRPC1 expression level in 32d-p210 cells. In addition to Orai1 channel, STIM1 can also directly activate TRPC1 [18, 54, 55]. Other works have reported that TRPC1 is involved in SOCE through its recruitment after STIM1-dependent Orai1 activation [21]. Since TRPC1 has been involved in SOCE in smooth [56] and skeletal [57, 58] muscles and in salivary gland [59], a decrease in TRPC1 level could also participate to the reduction of SOCE in Bcr-Abl-expressing progenitors. However, it is admitted that SOCE are mainly supported by STIM1-dependent Orai1 channels in hematopoietic cells such as lymphocytes [60].

Many studies deal with the link between SOCE and cancer. These studies propose that STIM1 and Orai1 mutations or deregulation of expression and/or activation of SOC constituents induced global modification of $Ca^{2+}$ entries leading to alteration of cancer homeostasis [23, 61, 62].

SOCE regulates the main cancer processes such as apoptosis, proliferation and migration/invasion in tumors from epithelial origin [27, 63, 64] through modification of $Ca^{2+}$ dependent pathways such as NFAT, Calmoduline (CaM) and its kinase (CaMK), and/or nuclear factor-kappa B (NFκB) (for review: [27, 65, 66]). As observed in several other models, SOCE activation induced NFAT1 and 2 nuclear translocation in both 32d cell lines (Figure 5). This is in agreement with the literature that links SOCE and NFAT pathway and suggests that NFAT could be a cancer target in solid tumors and leukemia (for review: [67]). Interestingly, NFAT translocation was reduced in 32d-p210 cells compared to 32dWT cells, a result to correlate with SOCE inhibition in Bcr-Abl-expressing cells. In addition, control conditions showed lower basal nuclear NFAT in 32d-p210 cells, suggesting that Bcr-Abl expression deregulates intracellular processes of $Ca^{2+}$ homeostasis and in turn $Ca^{2+}$-dependent signaling pathways such as the calcineurin/NFAT cascade. In fact, these reduced $Ca^{2+}$ influxes could inhibit NFAT pathway, a mechanism known to enhance myeloid differentiation characteristic of CML [68].

Interestingly, TRPC1 knockdown in endothelium progenitor cells induces SOCE decrease leading to a reduced proliferation and migration [69]. Here, TRPC1 expression and SOCE were reduced in 32d-p210 cells while proliferation rate remained higher than in control cells (Figures 3 and 5C). Nevertheless, SOCE seems to play a role in proliferation because SOCE inhibition induced a reduction of proliferation in both cell lines. This finding suggests that while SOCE regulates 32d cell proliferation, $Ca^{2+}$ independent pathways seem to be predominant in Bcr-Abl-dependent activation of enhanced cell proliferation. Bcr-Abl is known to control cell proliferation by tyrosine kinase dependent pathways such as Ras GTPase/ Mitogen-Activated Protein Kinase (MAPK), Phosphoinositide 3-kinase (PI3K)/ Protein kinase B (Akt) or Janus Kinase (JAK)/ Signal Transducers and Activators of Transcription (STAT) [70]. However, cell proliferation could be reduced in 32d-p210 cells by treatment with $Ca^{2+}$ channel blockers, showing that the inhibition of SOCE can help to counteract the proliferative effects of Bcr-Abl. This is in agreement with unpublished data from our laboratory indicating that cell proliferation is regulated by SOCE and tyrosine kinase activity through two distinct signaling cascades.

Migration and invasion are important for LSC development in CML, in part to evade hematopoietic niche control. Among other pathways, these mechanisms are also under the control of $Ca^{2+}$ influx. Bcr-Abl expression increased the length and speed of migration compared to control as already published (Figure 5E) [44, 45]. SOCE has been involved in migration in normal and cancer cells by regulating for instance cell adhesion and cytoskeleton dynamics [27]. The $Ca^{2+}$ signal could be local or global into cells and acts on $Ca^{2+}$ dependent proteins such as myosin light chain kinase (MLCK), PI3K or focal adhesion kinase (FAK) [46]. Migration is a key process for the Bcr-Abl stem cells to leave the hematopoetic niche to reach blood circulation where they proceed to final differentiation [71, 72]. Migration stimulation could reduce the recurrence of the disease by inducing the exit of the oncogenic cells from the niche. In this work, we showed that SOCE inhibition enhanced migration in both cell lines. SOCE inhibition in Bcr-Abl-expressing cells, which is linked to enhance migration, is in contradiction with most studies on other cell types. However, a recent publication showed a correlation between SOCE levels and invasiveness of different melanoma cells [48]. In this model, SOCE decrease was the consequence of Orai1 phosphorylation by PKC.

PKC has previously been involved in SOCE regulation by Orai1 phosphorylation [38, 73] and was also shown to regulate cell motility [74]. We were unsuccessful to show any modifications in Orai1 phosphorylation level in normal or SOCs-activated conditions (Figure 4C). Nevertheless, inhibition of PKC reduced SOCE in 32dWT cells, suggesting an enhancement of SOCE by the kinase in stimulated WT cells. The stimulating effect of PKC in 32dWT cells could be partly due to TRPC1, which confers PKC and $PIP_2$ activation on native TRPC-dependent SOCE in vascular myocytes [75, 76]. Interestingly, PKC inhibition had not effect on SOCE in Bcr-Abl-expressing cells (Figure 7A), supporting the idea that the PKC-dependent pathway enhancing SOCE in WT progenitor is inhibited by expression of Bcr-Abl, which in turn leads to the decrease in NFAT activation. The inhibition of PKC pathway could participate together with changes in Orai1 stoichiometry and decrease in TRPC1 level to the diminution of SOCE activation in Bcr-Abl-expressing cells. PKC phosphorylation regulates various pathways important in cancer development like proliferation, differentiation, migration and invasion. Different isoforms are known to be involved in different functions associated with CML as maturation of stem cells or differentiation. Depending on the isoform, the effect of Bcr-Abl on PKC is either activator (PKC iota, beta and alpha) or inhibitor (PKC delta). PKC modulation can interfere with Bcr-Abl-mediated deregulated pathways such as JAK/STAT, ERK or NFkB (for review: [49]). In combination with TKI, PKC activators (Bryostatin) have been tested in CML treatment showing a reduction of cell growth and increase of apoptosis [77, 78]. PKC is a good candidate to regulate SOCE in Bcr-Abl expressing cells even if no direct phosphorylation of SOC has been detected. A study using PEITC (Phenethylisothiocyanate), a potential cancer chemo preventive drug targeting PKC, described a crosstalk between Bcr-Abl and PKC though Raf1 and ERK1/2 pathways [79]. They showed that PEITC increased the sensibility to imatinib and reduced PKC expression. The double inhibitions of PKC and Bcr-Abl downregulated ERK4/2 and Raf1 pathways and these effects were not found in single inhibition.

Many alternative strategies are now developed in clinical trials to eradicate malignant TKI resistant clones. Among others, inhibition of signaling pathways (JAK2, PI3K/Akt/mTOR, Wnt/βcatenin, etc.), immunological approaches (Interferon-α (IFNα), vaccination, etc.) or niche targeting (stromal cytokines, stroma adhesion, etc.) are proposed in CML to reach stem cells [80–83]. Among numerous studies exploring alternate therapeutic approaches in CML, few are interested in the role of $Ca^{2+}$ in Bcr-Abl-expressing cells. A clinical trial showed that a combination of NFAT inhibitor cyclosporine with dasatinib facilitated leukemia cells elimination and improved survival of Bcr-Abl acute lymphoblastic leukemia [84]. Nevertheless, the phase 1b of this clinical trial concluded that this combined treatment may not be well tolerated [85]. Although, a recent proteomic analysis revealed that $Ca^{2+}$ homeostasis actors could be good targets in TKI-resistant CML [86]. Altogether, currently available SOCE inhibitors as GSK-7975A, YM-58483 or SKF-96365 could be more interesting molecules although work is in progress to improve their specificity [87, 88].

In conclusion, we unveiled a deregulation of $Ca^{2+}$ signaling in Bcr-Abl-expressing cells. Above others, Bcr-Abl seems to tightly regulate SOCE using different strategies: first, by modifying the expression levels of Orai1 and TRPC1, but not of STIM1, leading to a modification of the stoichiometry of the complex between STIM1 and Orai1 and/or TRPC1; second, through PKC pathway inhibition: in WT cells, SOCEs are clearly dependent on PKC activity while in Bcr-Abl expressing cells the dependency is lost. This suggests that Bcr-Abl already inhibits PKC activity or expression, which in turn regulates $Ca^{2+}$ homeostasis. The reduction of SOCE activation by Bcr-Abl has various important effects, notably, the decrease of $Ca^{2+}$ signal duration in response to agonist such as thrombin, and the lower activation of NFAT transcription factor, which needs to be dephosphorylated by the $Ca^{2+}$-dependent phosphatase calcineurin and is maintained into the nucleus by long-lasting $Ca^{2+}$ response. By decreasing SOCE activation, Bcr-Abl is thus able to reduce the activation of $Ca^{2+}$-dependent signaling pathways such as the calcineurin/NFAT cascade. Future studies aiming at enhancing SOCEs in CML treatment in parallel with TKI should open new perspectives.

# MATERIALS AND METHODS

## Antibodies and reagents

Antibodies against TRPC1 (SantaCruz), STIM1/GOK (BD Biosciences), NFAT1 and NFAT2 (Abcam) were mouse monoclonal. Antibodies against Orai1 (SantaCruz) and β-actin (Sigma-Aldrich) were rabbit polyclonal. Secondary horseradish peroxidase-conjugated (HRP)-conjugated goat anti-mouse IgG and goat anti-rabbit IgG were obtained from GE Healthcare (VWR International). Cyclopiazonic acid (CPA); *N,N,N′,N′*-tetrakis(2-pyridylmethyl)ethilenediamine (TPEN); Thrombin; 1{β-[3-(4-methoxyphenyl)propoxyl]-4-methoxyphenethyl}-

1*H*-imidazole hydrochloride (SKF-96365); 4-methyl-4′-[3,5-bis(trifluoromethyl)-1H-pyrazol-1-yl]-1,2,3-thiadiazole-5-carboxanilide (YM-58483); Cyclosporin A (CsA) and Bisindolylmaleimide (BIM) were all purchased from Sigma-Aldrich. Fura-2-Acetoxymethyl ester (Fura-2 AM) was acquired from SantaCruz, GSK-7579A from Aobious, INC; whereas 1,3-bis [bis (pyridin-2-ylméthyl) amino] propan-2-olato dizinc (II) complex (Phos-tag™ Acrylamide) was obtained from Wako Pure Chemical Industries Ltd. Imatinib mesylate was a gift from Dr E. Buchdunger (Novartis).

### Cell culture

32d is an Interleukine- 3 (IL-3)-dependent myeloid progenitor cell line of murine origin. 32d-p210 cells were generated as previously described [89, 90]. All cells were maintained in standard RPMI 1640 medium (Lonza AG) containing 10% FBS (Foetal Bovine Serum) (Gibco BRL; Fisher Scientific) and 50 µg/ml gentamycin (Sigma-Aldrich). Wild-type 32d (32dWT) culture medium was supplemented with 0.05 ng/ml recombinant mouse IL-3 (Miltenyi Biotech). All cells were maintained in a humidified 37° C incubator with 5% $CO_2$.

### Cytosolic $Ca^{2+}$ measurements

Cells were immobilized on fibronectin-coated coverslips for 4 hours and incubated in 1.8 mM $Ca^{2+}$ buffer (130 mM NaCl, 5.4 mM KCl, 1.8 mM $CaCl_2$, 0.8 mM $MgCl_2$, 10 mM HEPES, and 5.6 mM glucose, pH 7.4, with NaOH) and loaded with 3 µM Fura 2-AM for 30 minutes at 37° C, 5% $CO_2$. Next, cells were washed with 1.8 mM $Ca^{2+}$ solution. $Ca^{2+}$ influx were measured in response to 1 U/ml thrombin in 1.8 mM $Ca^{2+}$ buffer or after ER depletion induced by 15 µM CPA or 1 µM TPEN in $Ca^{2+}$-free solution (130 mM NaCl, 5.4 mM KCl, 0.1 mM EGTA, 0.8 mM $MgCl_2$, 10 mM HEPES, and 5.6 mM glucose, pH 7.4, with NaOH). Cells were treated or not with 10 µM imatinib for 4 hours or 10 µM BIM for 30 minutes. Fura 2-loaded cells were excited alternatively at 340 and 380 nm with a CAIRN monochromator (Cairn Research Limited, Faversham, UK), and the fluorescence emissions of each cell were simultaneously monitored at 510 nm using a CCD camera (Photonic Science, Robertsbridge, UK) coupled to an Olympus IX70 inverted microscope (×40 water immersion fluorescence objective). The variation of fluorescence was recorded with the Imaging Workbench 4.0 (IW 4.0) software (IndecBioSystems, Mountain View, CA, USA). Intracellular $Ca^{2+}$ measurements are shown as 340/380 nm ratios and analyzed with Origin 5.0 Software (Microcal Software, Northampton, MA, USA).

### Cell lysis and western blotting

Proteins were extracted using RIPA buffer (50 mM Tris·HCl pH 7.4, 150 mM NaCl, 5 mM EDTA, 5 mM $MgCl_2$, 0.05% NP-40, 1% sodium deoxycholate, 1% Triton X-100, 0.1% SDS supplemented with anti-phosphatase 1X and anti-protease cocktail 1X (Sigma-Aldrich)). Cell extract was then sonicated and heated during 5 minutes at 95° C. Protein extracts were dosed using BCA kit (Sigma-Aldrich). Samples were then subjected to SDS-PAGE and Western blot analysis as described previously [91, 92]. The blots were treated with primary antibody solutions overnight then for 1 hour with secondary HRP-conjugated antibody and developed with enhanced chemiluminescence kit (Merck Millipore). The apparent molecular weight was estimated according to the position of prestained protein markers (Kaleidoscope; Bio-Rad). The density of β-actin served as an internal loading control and levels of protein expression were quantified by GeneTools image analysis 4.3.5 Software.

### Phosphorylation assay

Cells were incubated in 1.8 mM $Ca^{2+}$ buffer (control) or in $Ca^{2+}$-free solution supplemented with 15 µM CPA for 6 minutes at room temperature (RT) and then stimulated with 1.8 mM $Ca^{2+}$ buffer for 1 minute before centrifugation. Proteins were extracted like described previously. 25 µM of acrylamide-pendant Phos-Tag™ ligand and 50 µM of $MnCl_2$ were included in the separating gel before polymerization according to the protocol of Kinoshita and al., [39]. Dephosphorylated sample was obtained in NE buffer 1X (50 mM Hepes, 100 mM NaCl, 2 mM DTT, 0.01% Brij 35, pH7.5 at 25° C) supplemented with 1 mM $MnCl_2$ and 1U Protein Phosphatase Lambda (New England Biolabs) during 1 hour at 30° C.

### Imaging of NFAT translocation

Cells were immobilized on fibronectin-coated coverslips and incubated in complete medium supplemented with 2 mM $CaCl_2$ and 10 mM Hepes (Sigma-Aldrich). Cells were stimulated for 7 minutes at 37° C with $Ca^{2+}$-free solution + 15 µM CPA or 30 minutes with 1.8 mM $Ca^{2+}$ buffer + 1 U/ml thrombin. After CPA stimulation, 1.8 mM $Ca^{2+}$ buffer + 15 µM CPA with or without 40 µM SKF-96365 or 10 µM YM-58483 were added during 30 minutes. 1 µM CsA was added on cells 30 minutes prior to CPA incubation. Cells were fixed in PBS (130 mM NaCl; 2.07 mM KCl; 1.5 mM $Na_2HPO_4$; 8 mM $KH_2PO_4$, pH 7.4)/4% paraformaldehyde for 30 minutes and permeabilized with PBS/0.3% Triton X-100 for 5 minutes. Primary antibodies were added overnight in PBS/1% BSA (dilution 1/50 for NFAT1 and 1/100 for NFTA2) and then with fluorescent secondary antibodies (1/500) for 1 hour. 1 µg/ml TOPRO and 1 µg/ml phalloidin (Sigma-Aldrich) were added for 20 minutes. Fluorescence was observed with confocal laser scanning microscopy using a Bio-Rad MRC 1024 ES equipped with an argon-krypton gas laser. Data were acquired using an inverted microscope (Olympus IX70) with ×60 oil immersion objective (image processing was performed with Olympus Fluoview (FV10-ASW 2.0)). Quantification of nuclear translocation was assessed by calculating the nuclear/cytosolic fluorescence intensity mean ratio of NFAT by Image J Software.

### Cell proliferation assay

Cells were cultured at a concentration of $2 \times 10^5$ cells/ml in complete medium and exposed to 10 μM YM-58483. Cell proliferation was assessed by counting on Malassez chamber at 24, 48 and 72 hours using trypan blue solution (Sigma-Aldrich).

### Cell cycle analysis assay

Cells were fixed in ice-cold ethanol (70% v/v) and stained with a solution of propidium iodide (PI) (5 μg/ml PI; 200 μg/ml RNase; 0.1% Triton X-100 in PBS) in the dark at RT for 1 hour 30 minutes. The DNA content was determined using a FACSVerse flow cytometer (BD Biosciences). Cell-cycle distribution was analysed using FlowJo 10.1 Software. Cells with DNA content between 2N and 4N were designated as being in the G1, S, or G2/M phase of the cell cycle. The number of cells in each compartment of the cell cycle was expressed as a percentage of the total number of counted cells (50.000 events).

### Migration assay

Cells were incubated on fibronectin coated coverslips and exposed to 40 μM SKF-96365. Cell migration was recorded during 30 minutes at 37° C with spinning disk confocal microscope (Olympus IX81-ZDC). Length and straightness of the migrating cells were measured with Imaris Software.

### Statistics

Experimental findings were expressed as mean ± standard deviation. Graphs were created using GraphPad Prism 5 Software. Statistical analysis was performed by Student's $t$ test. $P < 0.05^*$, $P < 0.01^{**}$ and $P < 0.001^{***}$.

### Author contributions

Hélène Cabanas, Bruno Constantin, Nicolas Bourmeyster and Nadine Déliot, for the conception and design of the work and wrote the manuscript with contributions from other authors. Hélène Cabanas, Thomas Harnois, Christophe Magaud, Laëtitia Cousin and Nadine Déliot performed experiments and/or data analysis. Bruno Constantin for critical revision of the manuscript. Nicolas Bourmeyster and Nadine Déliot, supervised the work.

### ACKNOWLEDGMENTS

The authors thank Dr E. Buchdunger (Novartis) for giving Imatinig mesylate. We are indebted to Anne Cantereau-Becq for her technical assistance with confocal microscopy and Adriana Delwail for her help with flow-cytometry. We thank Image UP core facilities.

### CONFLICTS OF INTEREST

The authors disclose no conflicts of interest.

### FUNDING

This work was funded by La Ligue Nationale Contre le Cancer Grand Ouest Comités de la Vienne et Charentes Maritimes, Région Poitou-Charentes, Cancéropôle Grand Ouest, and ACI from Universities of Tours and of Poitiers. Hélène Cabanas held a Ph.D. fellowship from the Region Poitou-Charentes.

### REFERENCES

1. Sawyers CL. Chronic myeloid leukemia. N Engl J Med. 1999; 340:1330–40. https://doi.org/10.1056/NEJM199904293401706.

2. Quintás-Cardama A, Cortes J. Molecular biology of bcr-abl1–positive chronic myeloid leukemia. Blood. 2009; 113:1619–30. https://doi.org/10.1182/blood-2008-03-144790.

3. Cortes J, Quintás-Cardama A, Kantarjian HM. Monitoring molecular response in chronic myeloid leukemia. Cancer. 2011; 117:1113–22. https://doi.org/10.1002/cncr.25527.

4. Rousselot P, Charbonnier A, Cony-Makhoul P, Agape P, Nicolini FE, Varet B, Gardembas M, Etienne G, Réa D, Roy L, Escoffre-Barbe M, Guerci-Bresler A, Tulliez M, et al. Loss of major molecular response as a trigger for restarting tyrosine kinase inhibitor therapy in patients with chronic-phase chronic myelogenous leukemia who have stopped imatinib after durable undetectable disease. J Clin Oncol. 2014; 32:424–30. https://doi.org/10.1200/JCO.2012.48.5797.

5. Druker BJ. Translation of the Philadelphia chromosome into therapy for CML. Blood. 2008; 112:4808–17. https://doi.org/10.1182/blood-2008-07-077958.

6. Berridge MJ. Calcium signalling remodelling and disease. Biochem Soc Trans. 2012; 40:297–309. https://doi.org/10.1042/BST20110766.

7. Ambudkar IS, Ong HL. Organization and function of TRPC channelosomes. Pflugers Arch. 2007; 455:187–200. https://doi.org/10.1007/s00424-007-0252-0.

8. Parekh AB, Putney JW. Store-operated calcium channels. Physiol Rev. 2005; 85:757–810. https://doi.org/10.1152/physrev.00057.2003.

9. Prakriya M, Lewis RS. Store-Operated Calcium Channels. Physiol Rev. 2015; 95:1383–436. https://doi.org/10.1152/physrev.00020.2014.

10. Cahalan MD. STIMulating store-operated Ca(2+) entry. Nat Cell Biol. 2009; 11:669–77. https://doi.org/10.1038/ncb0609-669.

11. Penna A, Demuro A, Yeromin AV, Zhang SL, Safrina O, Parker I, Cahalan MD. The CRAC channel consists of a tetramer formed by Stim-induced dimerization of Orai dimers. Nature. 2008; 456:116–20. https://doi.org/10.1038/nature07338.

12. Prakriya M, Feske S, Gwack Y, Srikanth S, Rao A, Hogan PG. Orai1 is an essential pore subunit of the CRAC channel. Nature. 2006; 443:230–3. https://doi.org/10.1038/nature05122.

13. Amcheslavsky A, Wood ML, Yeromin AV, Parker I, Freites JA, Tobias DJ, Cahalan MD. Molecular biophysics of Orai store-operated Ca2+ channels. Biophys J. 2015; 108:237–46. https://doi.org/10.1016/j.bpj.2014.11.3473.

14. Hogan PG, Rao A. Store-operated calcium entry: Mechanisms and modulation. Biochem Biophys Res Commun. 2015; 460:40–9. https://doi.org/10.1016/j.bbrc.2015.02.110.

15. Rothberg BS, Wang Y, Gill DL. Orai channel pore properties and gating by STIM: implications from the Orai crystal structure. Sci Signal. 2013; 6:pe9. https://doi.org/10.1126/scisignal.2003971.

16. Shim AH, Tirado-Lee L, Prakriya M. Structural and functional mechanisms of CRAC channel regulation. J Mol Biol. 2015; 427:77–93. https://doi.org/10.1016/j.jmb.2014.09.021.

17. Ong HL, de Souza LB, Ambudkar IS. Role of TRPC Channels in Store-Operated Calcium Entry. Adv Exp Med Biol. 2016; 898:87–109. https://doi.org/10.1007/978-3-319-26974-0_5.

18. Asanov A, Sampieri A, Moreno C, Pacheco J, Salgado A, Sherry R, Vaca L. Combined single channel and single molecule detection identifies subunit composition of STIM1-activated transient receptor potential canonical (TRPC) channels. Cell Calcium. 2015; 57:1–13. https://doi.org/10.1016/j.ceca.2014.10.011.

19. Worley PF, Zeng W, Huang GN, Yuan JP, Kim JY, Lee MG, Muallem S. TRPC channels as STIM1-regulated store-operated channels. Cell Calcium. 2007; 42:205–11. https://doi.org/10.1016/j.ceca.2007.03.004.

20. Cheng KT, Ong HL, Liu X, Ambudkar IS. Contribution and regulation of TRPC channels in store-operated Ca2+ entry. Curr Top Membr. 2013; 71:149–79. https://doi.org/10.1016/B978-0-12-407870-3.00007-X.

21. Cheng KT, Liu X, Ong HL, Swaim W, Ambudkar IS. Local Ca$^{2+}$ entry via Orai1 regulates plasma membrane recruitment of TRPC1 and controls cytosolic Ca$^{2+}$ signals required for specific cell functions. PLoS Biol. 2011; 9:e1001025. https://doi.org/10.1371/journal.pbio.1001025.

22. de Souza LB, Ong HL, Liu X, Ambudkar IS. Fast endocytic recycling determines TRPC1-STIM1 clustering in ER-PM junctions and plasma membrane function of the channel. Biochim Biophys Acta. 2015; 1853:2709–21. https://doi.org/10.1016/j.bbamcr.2015.07.019.

23. Dubois C, Vanden Abeele F, Lehen'kyi V, Gkika D, Guarmit B, Lepage G, Slomianny C, Borowiec AS, Bidaux G, Benahmed M, Shuba Y, Prevarskaya N. Remodeling of channel-forming ORAI proteins determines an oncogenic switch in prostate cancer. Cancer Cell. 2014; 26:19–32. https://doi.org/10.1016/j.ccr.2014.04.025.

24. Lee JM, Davis FM, Roberts-Thomson SJ, Monteith GR. Ion channels and transporters in cancer. 4. Remodeling of Ca(2+) signaling in tumorigenesis: role of Ca(2+) transport. Am J Physiol Cell Physiol. 2011; 301:C969–976. https://doi.org/10.1152/ajpcell.00136.2011.

25. Monteith GR, Davis FM, Roberts-Thomson SJ. Calcium channels and pumps in cancer: changes and consequences. J Biol Chem. 2012; 287:31666–73. https://doi.org/10.1074/jbc.R112.343061.

26. Capiod T. Cell proliferation, calcium influx and calcium channels. Biochimie. 2011; 93:2075–9. https://doi.org/10.1016/j.biochi.2011.07.015.

27. Déliot N, Constantin B. Plasma membrane calcium channels in cancer: Alterations and consequences for cell proliferation and migration. Biochim Biophys Acta. 2015; 1848:2512–22. https://doi.org/10.1016/j.bbamem.2015.06.009.

28. Fiorio Pla A, Kondratska K, Prevarskaya N. STIM and ORAI proteins: crucial roles in hallmarks of cancer. Am J Physiol Cell Physiol. 2016; 310:C509–519. https://doi.org/10.1152/ajpcell.00364.2015.

29. Hooper R, Zaidi MR, Soboloff J. The heterogeneity of store-operated calcium entry in melanoma. Sci China Life Sci. 2016; 59:764–9. https://doi.org/10.1007/s11427-016-5087-5.

30. Nielsen N, Lindemann O, Schwab A. TRP channels and STIM/ORAI proteins: sensors and effectors of cancer and stroma cell migration. Br J Pharmacol. 2014; 171:5524–40. https://doi.org/10.1111/bph.12721.

31. Piwocka K, Vejda S, Cotter TG, O'Sullivan GC, McKenna SL. Bcr-Abl reduces endoplasmic reticulum releasable calcium levels by a Bcl-2-independent mechanism and inhibits calcium-dependent apoptotic signaling. Blood. 2006; 107:4003–10. https://doi.org/10.1182/blood-2005-04-1523.

32. Vichalkovski A, Kotevic I, Gebhardt N, Kaderli R, Porzig H. Tyrosine kinase modulation of protein kinase C activity regulates G protein-linked Ca2+ signaling in leukemic hematopoietic cells. Cell Calcium. 2006; 39:517–28. https://doi.org/10.1016/j.ceca.2006.03.001.

33. Chantôme A, Potier-Cartereau M, Clarysse L, Fromont G, Marionneau-Lambot S, Guéguinou M, Pagès JC, Collin C, Oullier T, Girault A, Arbion F, Haelters JP, Jaffrès PA, et al. Pivotal role of the lipid Raft SK3-Orai1 complex in human cancer cell migration and bone metastases. Cancer Res. 2013; 73:4852–61. https://doi.org/10.1158/0008-5472.CAN-12-4572.

34. Hoover PJ, Lewis RS. Stoichiometric requirements for trapping and gating of Ca2+ release-activated Ca2+ (CRAC) channels by stromal interaction molecule 1 (STIM1). Proc Natl Acad Sci U S A. 2011; 108:13299–304. https://doi.org/10.1073/pnas.1101664108.

35. Li Z, Liu L, Deng Y, Ji W, Du W, Xu P, Chen L, Xu T. Graded activation of CRAC channel by binding of different numbers of STIM1 to Orai1 subunits. Cell Res. 2011; 21:305–15. https://doi.org/10.1038/cr.2010.131.

36. Ahmmed GU, Mehta D, Vogel S, Holinstat M, Paria BC, Tiruppathi C, Malik AB. Protein kinase Calpha phosphorylates the TRPC1 channel and regulates store-operated Ca2+ entry in endothelial cells. J Biol Chem. 2004; 279:20941–9. https://doi.org/10.1074/jbc.M313975200.

37. Saleh SN, Albert AP, Peppiatt-Wildman CM, Large WA. Diverse properties of store-operated TRPC channels activated by protein kinase C in vascular myocytes. J Physiol. 2008; 586:2463–76. https://doi.org/10.1113/jphysiol.2008.152157.

38. Kawasaki T, Ueyama T, Lange I, Feske S, Saito N. Protein kinase C-induced phosphorylation of Orai1 regulates the intracellular Ca2+ level via the store-operated Ca2+ channel. J Biol Chem. 2010; 285:25720–30. https://doi.org/10.1074/jbc.M109.022996.

39. Kinoshita E, Takahashi M, Takeda H, Shiro M, Koike T. Recognition of phosphate monoester dianion by an alkoxide-bridged dinuclear zinc(II) complex. Dalton Trans. 2004; 2004:1189–93. https://doi.org/10.1039/b400269e.

40. Lopez E, Jardin I, Berna-Erro A, Bermejo N, Salido GM, Sage SO, Rosado JA, Redondo PC. STIM1 tyrosine-phosphorylation is required for STIM1-Orai1 association in human platelets. Cell Signal. 2012; 24:1315–22. https://doi.org/10.1016/j.cellsig.2012.02.012.

41. Pozo-Guisado E, Martin-Romero FJ. The regulation of STIM1 by phosphorylation. Commun Integr Biol. 2013; 6:e26283. https://doi.org/10.4161/cib.26283.

42. Shou J, Jing J, Xie J, You L, Jing Z, Yao J, Han W, Pan H. Nuclear factor of activated T cells in cancer development and treatment. Cancer Lett. 2015; 361:174–84. https://doi.org/10.1016/j.canlet.2015.03.005.

43. O'Hare T, Deininger MW, Eide CA, Clackson T, Druker BJ. Targeting the BCR-ABL signaling pathway in therapy-resistant Philadelphia chromosome-positive leukemia. Clin Cancer Res. 2011; 17:212–21. https://doi.org/10.1158/1078-0432.CCR-09-3314.

44. Daubon T, Chasseriau J, El Ali A, Rivet J, Kitzis A, Constantin B, Bourmeyster N. Differential motility of p190bcr-abl- and p210bcr-abl-expressing cells: respective roles of Vav and Bcr-Abl GEFs. Oncogene. 2008; 27:2673–85. https://doi.org/10.1038/sj.onc.1210933.

45. Rochelle T, Daubon T, Van Troys M, Harnois T, Waterschoot D, Ampe C, Roy L, Bourmeyster N, Constantin B. p210bcr-abl induces amoeboid motility by recruiting ADF/destrin through RhoA/ROCK1. FASEB J. 2013; 27:123–34. https://doi.org/10.1096/fj.12-205112.

46. Chen YF, Hsu KF, Shen MR. The store-operated Ca(2+) entry-mediated signaling is important for cancer spread. Biochim Biophys Acta. 2016; 1863:1427–35. https://doi.org/10.1016/j.bbamcr.2015.11.030.

47. Gkika D, Flourakis M, Lemonnier L, Prevarskaya N. PSA reduces prostate cancer cell motility by stimulating TRPM8 activity and plasma membrane expression. Oncogene. 2010; 29:4611–6. https://doi.org/10.1038/onc.2010.210.

48. Hooper R, Zhang X, Webster M, Go C, Kedra J, Marchbank K, Gill DL, Weeraratna AT, Trebak M, Soboloff J. Novel Protein Kinase C-Mediated Control of Orai1 Function in Invasive Melanoma. Mol Cell Biol. 2015; 35:2790–8. https://doi.org/10.1128/MCB.01500-14.

49. Mencalha AL, Corrêa S, Abdelhay E. Role of calcium-dependent protein kinases in chronic myeloid leukemia: combined effects of PKC and BCR-ABL signaling on cellular alterations during leukemia development. Onco Targets Ther. 2014; 7:1247–54. https://doi.org/10.2147/OTT.S64303.

50. Harisseh R, Chatelier A, Magaud C, Déliot N, Constantin B. Involvement of TRPV2 and SOCE in calcium influx disorder in DMD primary human myotubes with a specific contribution of α1-syntrophin and PLC/PKC in SOCE regulation. Am J Physiol Cell Physiol. 2013; 304:C881–894. https://doi.org/10.1152/ajpcell.00182.2012.

51. Ciarcia R, d'Angelo D, Pacilio C, Pagnini D, Galdiero M, Fiorito F, Damiano S, Mattioli E, Lucchetti C, Florio S, Giordano A. Dysregulated calcium homeostasis and oxidative stress in chronic myeloid leukemia (CML) cells. J Cell Physiol. 2010; 224:443–53. https://doi.org/10.1002/jcp.22140.

52. Derler I, Jardin I, Romanin C. Molecular mechanisms of STIM/Orai communication. Am J Physiol Cell Physiol. 2016; 310:C643–662. https://doi.org/10.1152/ajpcell.00007.2016.

53. Hou X, Pedi L, Diver MM, Long SB. Crystal structure of the calcium release-activated calcium channel Orai. Science. 2012; 338:1308–13. https://doi.org/10.1126/science.1228757.

54. Huang GN, Zeng W, Kim JY, Yuan JP, Han L, Muallem S, Worley PF. STIM1 carboxyl-terminus activates native SOC, I(crac) and TRPC1 channels. Nat Cell Biol. 2006; 8:1003–10. https://doi.org/10.1038/ncb1454.

55. Lee KP, Choi S, Hong JH, Ahuja M, Graham S, Ma R, So I, Shin DM, Muallem S, Yuan JP. Molecular determinants mediating gating of Transient Receptor Potential Canonical (TRPC) channels by stromal interaction molecule 1 (STIM1). J Biol Chem. 2014; 289:6372–82. https://doi.org/10.1074/jbc.M113.546556.

56. Gonzalez-Cobos JC, Trebak M. TRPC channels in smooth muscle cells. Front Biosci. 2010; 15:1023–39.
57. Sabourin J, Lamiche C, Vandebrouck A, Magaud C, Rivet J, Cognard C, Bourmeyster N, Constantin B. Regulation of TRPC1 and TRPC4 cation channels requires an alpha1-syntrophin-dependent complex in skeletal mouse myotubes. J Biol Chem. 2009; 284:36248–61. https://doi.org/10.1074/jbc.M109.012872.
58. Zanou N, Shapovalov G, Louis M, Tajeddine N, Gallo C, Van Schoor M, Anguish I, Cao ML, Schakman O, Dietrich A, Lebacq J, Ruegg U, Roulet E, et al. Role of TRPC1 channel in skeletal muscle function. Am J Physiol Cell Physiol. 2010; 298:C149–162. https://doi.org/10.1152/ajpcell.00241.2009.
59. Cheng KT, Liu X, Ong HL, Ambudkar IS. Functional requirement for Orai1 in store-operated TRPC1-STIM1 channels. J Biol Chem. 2008; 283:12935–40. https://doi.org/10.1074/jbc.C800008200.
60. Shaw PJ, Feske S. Regulation of lymphocyte function by ORAI and STIM proteins in infection and autoimmunity. J Physiol. 2012; 590:4157–67. https://doi.org/10.1113/jphysiol.2012.233221.
61. Jardin I, Rosado JA. STIM and calcium channel complexes in cancer. Biochim Biophys Acta. 2016; 1863:1418–26. https://doi.org/10.1016/j.bbamcr.2015.10.003.
62. Lacruz RS, Feske S. Diseases caused by mutations in ORAI1 and STIM1. Ann N Y Acad Sci. 2015; 1356:45–79. https://doi.org/10.1111/nyas.12938.
63. Mognol GP, Carneiro FR, Robbs BK, Faget DV, Viola JP. Cell cycle and apoptosis regulation by NFAT transcription factors: new roles for an old player. Cell Death Dis. 2016; 7:e2199. https://doi.org/10.1038/cddis.2016.97.
64. Qin JJ, Nag S, Wang W, Zhou J, Zhang WD, Wang H, Zhang R. NFAT as cancer target: mission possible? Biochim Biophys Acta. 2014; 1846:297–311. https://doi.org/10.1016/j.bbcan.2014.07.009.
65. Pinto MC, Kihara AH, Goulart VA, Tonelli FM, Gomes KN, Ulrich H, Resende RR. Calcium signaling and cell proliferation. Cell Signal. 2015; 27:2139–49. https://doi.org/10.1016/j.cellsig.2015.08.006.
66. Rinkenbaugh AL, Baldwin AS. The NF-κB Pathway and Cancer Stem Cells. Cells. 2016; 5. https://doi.org/10.3390/cells5020016.
67. Medyouf H, Ghysdael J. The calcineurin/NFAT signaling pathway: a novel therapeutic target in leukemia and solid tumors. Cell Cycle. 2008; 7:297–303. https://doi.org/10.4161/cc.7.3.5357.
68. Kiani A, Kuithan H, Kuithan F, Kyttälä S, Habermann I, Temme A, Bornhäuser M, Ehninger G. Expression analysis of nuclear factor of activated T cells (NFAT) during myeloid differentiation of CD34+ cells: regulation of Fas ligand gene expression in megakaryocytes. Exp Hematol. 2007; 35:757–70.
69. Kuang CY, Yu Y, Wang K, Qian DH, Den MY, Huang L. Knockdown of transient receptor potential canonical-1 reduces the proliferation and migration of endothelial progenitor cells. Stem Cells Dev. 2012; 21:487–96. https://doi.org/10.1089/scd.2011.0027.
70. Warsch W, Walz C, Sexl V. JAK of all trades: JAK2-STAT5 as novel therapeutic targets in BCR-ABL1+ chronic myeloid leukemia. Blood. 2013; 122:2167–75. https://doi.org/10.1182/blood-2013-02-485573.
71. Sloma I, Jiang X, Eaves AC, Eaves CJ. Insights into the stem cells of chronic myeloid leukemia. Leukemia. 2010; 24:1823–33. https://doi.org/10.1038/leu.2010.159.
72. Wang X, Huang S, Chen JL. Understanding of leukemic stem cells and their clinical implications. Mol Cancer. 2017; 16:2. https://doi.org/10.1186/s12943-016-0574-7.
73. Parekh AB, Penner R. Depletion-activated calcium current is inhibited by protein kinase in RBL-2H3 cells. Proc Natl Acad Sci U S A. 1995; 92:7907–11.
74. Long A, Freeley M. Protein kinase C: a regulator of cytoskeleton remodelling and T-cell migration. Biochem Soc Trans. 2014; 42:1490–7. https://doi.org/10.1042/BST20140204.
75. Saleh SN, Albert AP, Large WA. Activation of native TRPC1/C5/C6 channels by endothelin-1 is mediated by both PIP3 and PIP2 in rabbit coronary artery myocytes. J Physiol. 2009; 587:5361–75. https://doi.org/10.1113/jphysiol.2009.180331.
76. Shi J, Ju M, Abramowitz J, Large WA, Birnbaumer L, Albert AP. TRPC1 proteins confer PKC and phosphoinositol activation on native heteromeric TRPC1/C5 channels in vascular smooth muscle: comparative study of wild-type and TRPC1–/– mice. FASEB J. 2012; 2 6:409–19. https://doi.org/10.1096/fj.11-185611.
77. Lothstein L, Savranskaya L, Sweatman TW. N-Benzyladriamycin-14-valerate (AD 198) cytotoxicty circumvents Bcr-Abl anti-apoptotic signaling in human leukemia cells and also potentiates imatinib cytotoxicity. Leuk Res. 2007; 31:1085–95. https://doi.org/10.1016/j.leukres.2006.11.003.
78. Szallasi Z, Denning MF, Smith CB, Dlugosz AA, Yuspa SH, Pettit GR, Blumberg PM. Bryostatin 1 protects protein kinase C-delta from down-regulation in mouse keratinocytes in parallel with its inhibition of phorbol ester-induced differentiation. Mol Pharmacol. 1994; 46:840–50.
79. Roy M, Sarkar R, Mukherjee A, Mukherjee S. Inhibition of crosstalk between Bcr-Abl and PKC signaling by PEITC, augments imatinib sensitivity in chronic myelogenous leukemia cells. Chem Biol Interact. 2015; 242:195–201. https://doi.org/10.1016/j.cbi.2015.10.004.
80. Ciarcia R, Damiano S, Montagnaro S, Pagnini U, Ruocco A, Caparrotti G, d'Angelo D, Boffo S, Morales F, Rizzolio F, Florio S, Giordano A. Combined effects of PI3K and SRC kinase inhibitors with imatinib on intracellular calcium levels, autophagy, and apoptosis in CML-PBL cells. Cell Cycle. 2013; 12:2839–48. https://doi.org/10.4161/cc.25920.

81. Ahmed W, Van Etten RA. Alternative approaches to eradicating the malignant clone in chronic myeloid leukemia: tyrosine-kinase inhibitor combinations and beyond. Hematology Am Soc Hematol Educ Program. 2013; 2013:189–200. https://doi.org/10.1182/asheducation-2013.1.189.
82. Eide CA, O'Hare T. Chronic myeloid leukemia: advances in understanding disease biology and mechanisms of resistance to tyrosine kinase inhibitors. Curr Hematol Malig Rep. 2015; 10:158–66. https://doi.org/10.1007/s11899-015-0248-3.
83. Hamad A, Sahli Z, El Sabban M, Mouteirik M, Nasr R. Emerging therapeutic strategies for targeting chronic myeloid leukemia stem cells. Stem Cells Int. 2013; 2013:724360. https://doi.org/10.1155/2013/724360.
84. Gregory MA, Phang TL, Neviani P, Alvarez-Calderon F, Eide CA, O'Hare T, Zaberezhnyy V, Williams RT, Druker BJ, Perrotti D, Degregori J. Wnt/Ca2+/NFAT signaling maintains survival of Ph+ leukemia cells upon inhibition of Bcr-Abl. Cancer Cell. 2010; 18:74–87. https://doi.org/10.1016/j.ccr.2010.04.025.
85. Gardner LA, Klawitter J, Gregory MA, Zaberezhnyy V, Baturin D, Pollyea DA, Takebe N, Christians U, Gore L, DeGregori J, Porter CC. Inhibition of calcineurin combined with dasatinib has direct and indirect anti-leukemia effects against BCR-ABL1(+) leukemia. Am J Hematol. 2014; 89:896–903. https://doi.org/10.1002/ajh.23776.
86. Toman O, Kabickova T, Vit O, Fiser R, Polakova KM, Zach J, Linhartova J, Vyoral D, Petrak J. Proteomic analysis of imatinib-resistant CML-T1 cells reveals calcium homeostasis as a potential therapeutic target. Oncol Rep. 2016; 36:1258–68. https://doi.org/10.3892/or.2016.4945.
87. Jairaman A, Prakriya M. Molecular pharmacology of store-operated CRAC channels. Channels (Austin). 2013; 7:402–14. https://doi.org/10.4161/chan.25292.
88. Xu SZ. Assessing TRPC Channel Function Using Pore-Blocking Antibodies. In: Zhu MX, editor. TRP Channels: CRC Press/Taylor & Francis; 2011. Available from http://www.ncbi.nlm.nih.gov/books/NBK92829/.
89. Daley GQ, Baltimore D. Transformation of an interleukin 3-dependent hematopoietic cell line by the chronic myelogenous leukemia-specific P210bcr/abl protein. Proc Natl Acad Sci U S A. 1988; 85:9312–6.
90. Matulonis U, Salgia R, Okuda K, Druker B, Griffin JD. Interleukin-3 and p210 BCR/ABL activate both unique and overlapping pathways of signal transduction in a factor-dependent myeloid cell line. Exp Hematol. 1993; 21:1460–6.
91. Harnois T, Constantin B, Rioux A, Grenioux E, Kitzis A, Bourmeyster N. Differential interaction and activation of Rho family GTPases by p210bcr-abl and p190bcr-abl. Oncogene. 2003; 22:6445–54. https://doi.org/10.1038/sj.onc.1206626.
92. Laemmli UK, Favre M. Maturation of the head of bacteriophage T4. I. DNA packaging events. J Mol Biol. 1973; 80:575–99.